\documentclass[11pt,a4paper]{article}
\pdfoutput=1
\usepackage{jheppub}
\usepackage{amsfonts, amssymb, amsmath}
\usepackage{mathrsfs}
\usepackage{graphicx}
\usepackage[utf8]{inputenc}
\usepackage[russian,english]{babel}

\usepackage{color}
\definecolor{dark-gray}{gray}{0.20}
\definecolor{gray}{gray}{0.30}
\definecolor{light-gray}{gray}{0.80}
\definecolor{dark-red}{rgb}{0.7,0,0}
\definecolor{dark-green}{rgb}{0.1,0.4,0}
\definecolor{dark-blue}{rgb}{0.3,0.3,0.7}
\definecolor{light-blue}{rgb}{0.8,0.8,1}
\definecolor{blue}{rgb}{0,0,1}
\definecolor{red}{rgb}{1,0,0}
\definecolor{green}{rgb}{0,1,0}

\usepackage{hyperref}
\hypersetup{
	colorlinks=true,
	linkcolor=blue,
	citecolor=blue,
	urlcolor=blue,
	linktoc=page
}

\def\cB{{\cal B}}

\def\cF{{\cal F}}

\def\cN{{\cal N}}

\newcommand{\be}{\begin{equation}}
\newcommand{\ee}{\end{equation}}
\newcommand{\bea}{\begin{eqnarray}}
\newcommand{\eea}{\end{eqnarray}}

\title{Spindle black holes in AdS$_4 \times$SE$_7$}

\author[a]{Kiril Hristov}
\author[b]{and Minwoo Suh}

\affiliation[a]{Faculty of Physics, Sofia University, J. Bourchier Blvd. 5, 1164 Sofia, Bulgaria}
\affiliation[a]{INRNE, Bulgarian Academy of Sciences, Tsarigradsko Chaussee 72, 1784 Sofia, Bulgaria}
\affiliation[b]{Kavli Institute for Theoretical Sciences, University of Chinese Academy of Sciences, \\ Beijing 100190, China}

\emailAdd{khristov@phys.uni-sofia.bg}
\emailAdd{minwoosuh1@gmail.com}

\abstract{
\noindent We construct new classes of supersymmetric AdS$_2 \times {{\Sigma}}$ solutions of 4d gauged supergravity in presence of charged hypermultiplet scalars, with ${{\Sigma}}$ the complex weighted projective space known as a {\it spindle}. These solutions can be viewed as near-horizon geometries of asymptotically Anti de-Sitter (AdS$_4$) black holes with magnetic fluxes that admit embedding in 11d on Sasaki-Einstein (SE$_7$) manifolds, which renders them of holographic interest. We show that in each case the Bekenstein-Hawking entropy follows from the procedure of gluing two gravitational blocks, ultimately determined by SE$_7$ data. This allows us to establish the general form of the gravitational blocks in gauged 4d $\cN=2$ supergravity with charged scalars and massive vectors. Holographically, our results provide a large N answer for the spindle index with anti-twist and additional mesonic or baryonic fluxes of a number of ${\mathcal{N}=2}$ Chern-Simons-matter theories.
}
\date{\today}
\begin{document}
\maketitle


\section{Introduction and main results}
\label{sec:intro}

The study of supersymmetric AdS solutions arising as brane constructions in string/M-theory and the parallel progress in exact gauge theory calculations on curved manifolds have greatly increased the detailed understanding of the holographic duality, \cite{Maldacena:1997re}. Recent research efforts have been focused on the possibility of branes wrapping orbifolds such as the complex weighted projective space $\Sigma = \mathbb{WCP}^1_{n_-, n_+}$,  or spindle, specified by two co-prime positive integers $n_\pm$. The first solutions of the type AdS$_3 \times \Sigma$ were analysed in 5d supergravity and interpreted in IIB theory as the near-horizon of D3 branes wrapping the spindle, \cite{Ferrero:2020laf}. Soon after, accelerating black holes in AdS$_4$ were shown to exhibit spindle horizons, corresponding to a full 11d solution of M2 branes wrapping spindles, \cite{Ferrero:2020twa}. Various generalizations followed these initial constructions, such as \cite{Hosseini:2021fge, Boido:2021szx, Ferrero:2021wvk, Cassani:2021dwa, Ferrero:2021ovq, Couzens:2021rlk, Faedo:2021nub, Ferrero:2021etw, Giri:2021xta, Couzens:2021cpk, Cheung:2022ilc, Suh:2022olh, Arav:2022lzo, Couzens:2022yiv, Couzens:2022aki, Couzens:2022lvg, Faedo:2022rqx, Suh:2022pkg,Suh:2023xse,Amariti:2023mpg, Kim:2023ncn} that were all focused on spindles or related disks, \cite{Bah:2021mzw, Bah:2021hei, Couzens:2021tnv, Suh:2021ifj, Suh:2021aik, Suh:2021hef, Karndumri:2022wpu, Couzens:2022yjl, Bah:2022yjf}, in truncations of the maximal supergravities admitting AdS vacuum, and their field theory duals. See also \cite{Gutperle:2022pgw, Gutperle:2023yrd}.

It is also natural to start looking at theories of less supersymmetry, allowing for more general classes of internal manifolds. The lower-dimensional supergravity truncations in these cases feature the addition of charged hypermultiplet scalars and massive vectors, rendering the analysis of the BPS solutions technically more challenging.  Still, the first example of such solutions with magnetic fluxes appeared in \cite{Arav:2022lzo} and they have been generalized to other models in \cite{Suh:2022pkg,Suh:2023xse,Amariti:2023mpg}. In the present work we focus on completing this task in the 4d setting, considering the consistent truncations of \cite{Cassani:2012pj} to $\cN=2$ supergravity with AdS$_4$ vacuum on homogeneous SE$_7$ spaces. More precisely, we focus on the cases of the SE$_7$ spaces $Q^{1,1,1}$ and $M^{1,1,1}$ which in practice contain the spindle solutions for all other homogeneous spaces in \cite{Cassani:2012pj} as discussed below. The spindle black holes we discover have vanishing electric charges and angular momentum and share some features with the spherical BPS black holes in these models considered in \cite{Halmagyi:2013sla}, but they no longer allow for an analytic form of the solution even for the near-horizon geometry. Our analysis is very similar to \cite{Suh:2022pkg}, where spindle horizons were studied in the $\cN=2$ AdS$_4 \times$S$^7$ vacuum of 11d, corresponding to a massive deformation of ABJM theory, \cite{Aharony:2008ug}, (mABJM). We also come back to it in view of the gravitational blocks we discuss next. Note that we limit ourselves to analysis of the near-horizon region and do not discuss the complete flow toward the asymptotic AdS$_4$ vacuum that is also only possible numerically,  \cite{Halmagyi:2013sla}.~\footnote{See also \cite{Kim:2020qec} for uplifts and \cite{Monten:2016tpu,Monten:2021som} for numeric solutions of thermal black holes in these theories.} We expect that the horizons we construct here are part of black hole geometries with {\it non-vanishing} acceleration parameter as in \cite{Ferrero:2020twa}.

A related recent development is the construction of gravitational blocks in \cite{Hosseini:2019iad}, where it was shown that the on-shell action and entropy of the BPS black holes can be recovered from simpler building blocks defined by supergravity data. Although initially constructed for various black holes with regular spherical horizons in AdS$_4$ and AdS$_5$, \cite{Maldacena:2000mw,Gutowski:2004yv,Cacciatori:2009iz,Hristov:2018spe,Hristov:2019mqp,Hosseini:2019lkt}, there is evidence that the gravitational blocks can be used to derive the on-shell action of {\it all} BPS backgrounds with fixed points, \cite{BenettiGenolini:2019jdz,Hristov:2021qsw,Hristov:2022plc,BenettiGenolini:2023kxp,Martelli:2023oqk}. Consequently, this logic was successfully applied to spindly constructions of various types and dimensions in \cite{Hosseini:2021fge,Faedo:2021nub,Faedo:2022rqx,Boido:2022iye,Boido:2022mbe,Suh:2023xse,Amariti:2023mpg}. In our present analysis we utilize the gravitational block picture in order to bring more transparency into the structure of the solutions we discover, since we find that these basic building blocks provide an analytic description of BPS equations that require numerical integration. In the same time, our results allow us to uncover the gravitational block construction for theories with charged hypermultiplet scalars and corresponding massive vectors, which was so far only considered for simpler solutions in \cite{Hosseini:2020vgl,Hosseini:2020wag}. 

The main logic behind constructing the gravitational blocks in presence of abelian charged hypermultiplets is in its essence rather straightforward. In the language of 4d $\cN=2$ gauged supergravity, the theory in the presence of $n_V$ vector multiplets and $n_H$ hypermultiplets is defined by the respective scalar manifolds in the two sectors and the choice of their symmetries (in the abelian case only of the hypermultiplet scalar manifold) to be gauged. There are $(n_V+1)$ $U(1)$ fundamental gauge fields $A^I$, $I = 0, 1, .. n_V$ that can be used to gauge the isometries (in a priori arbitrary linear combinations) and thus charge the corresponding scalars. Via supersymmetry, it turns out that this process leads to gauging the R-symmetry of the theory, i.e.\ the gravitini also become charged under another particular linear combination of the $U(1)$ vectors. One then has a number of ``massive'' vectors (labeled here by index $\alpha$) that appear in scalar covariant derivatives, that we can denote generally~\footnote{Note that here we only discuss the so-called {\it electric} gauging, i.e.\ we only use the fundamental gauge fields and not their duals. The latter correspond to magnetic gauging and are generally allowed in supergravity. It turns out all the models we consider here allow for a symplectic frame where the gauging can be purely electric, see \cite{Halmagyi:2013sla}. We give more comments about the general dyonic gauging in the discussion section.} as $A^m_\alpha = \zeta_{\alpha, I} A^I$, and the R-symmetry vector $A^R = \xi_I A^I$, where the coefficients $\zeta_{\alpha, I}$ and $\xi_I$ are in general functions of the hypermultiplet scalars. The BPS conditions set the corresponding scalars to particular constants (in a model-dependent way that we later discuss explicitly) such that we can consider $\zeta_{\alpha, I}$ and $\xi_I$ to be a set of constants that are uniquely fixed by the details of the hypermultiplet sector. On the other hand, the vector multiplet scalar manifold is defined by the so-called prepotential, $F(X^I)$, a homogeneous function of degree 2 of the sections $X^I$ that determine the complex scalars in a unique way.

As shown in \cite{Hosseini:2021fge,Faedo:2021nub}, the gravitational block construction of the on-shell action of black holes with spindle horizons (defined by the co-prime integers $n_-, n_+$) is simply given by~\footnote{See section \ref{sec:6} for the definition of a single gravitational block and its relation with the holographic free energy on the three-sphere.}
\be
\label{eq:1}	
I^\sigma (\varphi, \epsilon; n_\pm) =  \tfrac{i\, \pi}{8\, G^{(4)}_N}\,  \frac1{\epsilon} \left( F (\varphi^I +\epsilon\, \frak{n}^I) - \sigma F (\varphi^I -\epsilon\, \frak{n}^I) \right)\ ,
\ee
where $G_N^{(4)}$ the Newton constant and $\frak{n}^I$ the respective magnetic fluxes through the spindle. Supersymmetry further dictates that
\be
\label{eq:fluxes}
	\frak{n}^R = \xi_I \frak{n}^I = \frac{n_++\sigma{n}_-}{n_+n_-}\,, \qquad \frak{n}^m_\alpha = \zeta_{\alpha, I} \frak{n}^I = 0\ ,
\ee
where $\sigma = \pm 1$ reflects the Killing spinor orientation at the two poles of the spindle and is called twist and anti-twist, respectively. The black hole entropy function is then given as a Legendre transform of the on-shell action,
\be
\label{eq:2}
	S^\sigma (q, J, \varphi, \epsilon; n_\pm ) = \tfrac{i\, \pi}{2\, G^{(4)}_N} ( \varphi^I\, q_I + \epsilon\, J) - I^\sigma (\varphi, \epsilon; n_\pm) + \lambda (\xi_I \varphi^I - \tfrac{n_+ -\sigma\, n_-}{n_+ n_-}\, \epsilon - 2 ) + \mu^\alpha\, \zeta_{\alpha, I} \varphi^I\ ,
\ee
where, in order to recover the entropy in terms of the conserved charges, one needs to extremize the above functional with respect to the fugacities $\varphi^I$ and $\epsilon$ conjugate respectively to the conserved electric charges $q_I$ and angular momentum $J$,
\be
\label{eq:3}
	S^\sigma_\text{on-shell} (q, J; n_\pm ) = S^\sigma (q_I, J, \bar \varphi^I, \bar \epsilon; n_\pm ) \ , \qquad \partial_{\varphi^I, \epsilon}  S^\sigma |_{\bar{\varphi}^I, \bar{\epsilon}} = \partial_{\lambda, \mu^\alpha} S^\sigma |_{\bar{\varphi}^I, \bar{\epsilon} } = 0\ .  
\ee
Note that this is a constrained extremization due to the appearance of the Lagrange multipliers $\lambda, \mu^\alpha$. Apart from matching the on-shell Bekestein-Hawking entropy of the solutions we discover, we also show that the values of the vector multiplet scalars at the poles of the spindle are precisely related to the extremal values $\bar \varphi^I$ and $\bar \epsilon$.~\footnote{ Interestingly, the limit $n_- = n_+ = 1$ is smooth and recovers the results for spherical horizons, \cite{Hosseini:2019iad}, where again both the entropy and the scalars at the horizon can be recovered via extremization.} The novel feature of hypermultiplet gauging is that each massive multiplet contributes with an extra constraint ($\zeta_{\alpha, I} \varphi^I = 0$) effectively enforcing the decrease of the number of flavour symmetries, i.e.\ unconstrained $U(1)$ vectors. These constraints can be understood from the supersymmetry-preserving Higgs mechanism that takes place at the poles of the spindle, \cite{Hristov:2010eu,Hosseini:2017fjo}.

Finally, let us mention that the solutions we discuss here are holographically dual to 3d gauge theories, see \cite{Benini:2009qs,Cremonesi:2010ae}, on S$^1 \times \Sigma$. Recently their partition functions, called spindle indices, were defined for both the twist and the anti-twist choices above, $\sigma = \pm 1$, \cite{Inglese:2023wky}. We expect further work to reveal the large N expressions of the dual theories we consider here, and so the present results should be immediately comparable similarly to the spherical  black holes, \cite{Benini:2015noa, Benini:2015eyy}. We should mention that the gravitational block description allows for a general discussion of both signs for $\sigma$ but so far we have only found fully consistent solutions only in the anti-twist class. We also stress that the $\cN=2$ supergravity language used above does not distinguish between different types of abelian vector multiplets, which can be considered as flavour symmetries in addition to the R-symmetry. However, from a higher-dimensional point of view and holographic standpoint, these flavour symmetries can be divided into two main classes, called mesonic and baryonic following the definition in \cite{Hosseini:2019ddy}. We are going to see that our analysis features mesonic symmetries in the mABJM case, which are immediately translatable in field theory, while the $Q^{1,1,1}$ and $M^{1,1,1}$ cases exhibit only baryonic symmetries which are at present not properly understood holographically.\footnote{See e.g.\ section 4.4 of \cite{Azzurli:2017kxo} for a concise and clear discussion on this issue.} These features are entirely due to the available four-dimensional supergravity truncations. An alternative possibility, discussed towards the end, was put forward in \cite{Couzens:2018wnk,Hosseini:2019ddy,Gauntlett:2019roi,Kim:2019umc,Boido:2022iye,Boido:2022mbe} that consider the gravitational blocks directly in 10/11d.  

The rest of this paper is organized as follows. In section \ref{sec:2} (and more technically in App.\ \ref{appA}) we elaborate on the supergravity theory. In section \ref{sec:3} we write down the ansatz for background solutions and the corresponding BPS equations (with more details in App.\ \ref{appB}). In section \ref{sec:4} we discuss the explicit solutions, providing analytic results in the minimal truncation and numeric data for the more general solutions. In section \ref{sec:5} we discuss more briefly the case of $M^{1,1,1}$, which relates straightforwardly to the previous solutions. In section \ref{sec:6}, which can also be read in isolation from the rest, we discuss in detail the construction of gravitational blocks and their match with the explicit solutions. We finish the main body of this work with a list of open questions in section \ref{sec:7}. In addition, for the benefit of the interested reader, we have included a complementary {\it Mathematica} notebook with the present submission, containing details on the numerical solutions and gravitational block matching and allowing one to change explicitly the various solution parameters and magnetic fluxes.

\section{The supergravity model}
\label{sec:2}

We consider gauged $\mathcal{N}=2$ supergravity obtained from the dimensional reduction of eleven-dimensional supergravity on $Q^{1,1,1}$ manifold with 3 vector multiplets and a prepotential given by
\be
\label{eq:prepotential}
	F = -2 i\, \sqrt{X^0 X^1 X^2 X^3}\ ,
\ee 
and further details on the hypermultiplet gauging presented in appendix \ref{appA}. For the purposes of finding spindle horizons with magnetic fluxes only, in the appendix we perform a further truncation setting the axionic part of the vector multiplet scalars and two of the hypermultiplet scalars to zero. The remaining bosonic field content we consider here is the metric, four $U(1)$ gauge fields, $A^I$, $I=0,\ldots,3$, three real scalars from the vector multiplets, $u_i$, $i=1,2,3$ and two real scalars from the so-called universal hypermultiplet, $(\phi,\sigma)$. We have mostly plus signature. The bosonic Lagrangian of the truncation is
\begin{align} \label{lagmain}
e^{-1}\mathcal{L}\,&=\,\frac{1}{2}R-\sum_{i=1}^3\partial_\mu{}u_i\partial^\mu{}u_i-\partial_\mu\phi\partial^\mu\phi-\frac{1}{4}e^{4\phi}D_\mu\sigma{}D^\mu\sigma-g^2\mathcal{V} \notag \\
&-\Big[e^{2u_1+2u_2+2u_3}F^0_{\mu\nu}F^{0\mu\nu}+e^{2u_1-2u_2-2u_3}F^1_{\mu\nu}F^{1\mu\nu} \notag \\
& \,\,\,\, +e^{-2u_1+2u_2-2u_3}F^2_{\mu\nu}F^{2\mu\nu}+e^{-2u_1-2u_2+2u_3}F^3_{\mu\nu}F^{3\mu\nu}\Big]\,,
\end{align}
where
\begin{equation}
D\sigma\,=\,d\sigma+g\left(e_0A^0-2A^1-2A^2-2A^3\right)\,,
\end{equation}
and $e_0 > 0$ is an arbitrary Freund-Rubin parameter. The scalar potential is given by
\begin{equation}
\mathcal{V}\,=\,\sum_{i=1}^3\left(\frac{\partial\,W}{\partial\,u_i}\right)^2+\left(\frac{\partial\,W}{\partial\phi}\right)^2-3W^2\,,
\end{equation}
where the superpotential is
\begin{equation} \label{superpot}
W\,=\,\frac{1}{4}e^{-u_1-u_2-u_3}\left(8-e_0e^{2\phi}+2e^{2\phi}\left(e^{2u_1+2u_2}+e^{2u_2+2u_3}+e^{2u_3+2u_1}\right)\right)\,.
\end{equation}
The R-symmetry vector field, the massive vector field, and the two Betti vector fields are given by, respectively,
\begin{align} \label{vectorsrmbb}
A^{\text{R}}\,=&\,4\sqrt{2}A^0\,, \notag \\
A^{m}\,=&\,\sqrt{2}\left(e_0A^0-2A^1-2A^2-2A^3\right)\,, \notag \\
A^{\mathcal{B}_1}\,=&\, 4\sqrt{2}\left(A^1-2A^2+A^3\right)\,, \notag \\
A^{\mathcal{B}_2}\,=&\, 4\sqrt{2}\left(A^1-A^3\right)\,.
\end{align}

We present the supersymmetry variations of fermionic fields. The gravitino, gaugino and hyperino variations reduce to, respectively,~\footnote{Here we directly use Dirac spinors $\epsilon$, which can be constructed from the Weyl spinors $\varepsilon_A$ in the standard $\cN=2$ supergravity conventions, see appendix \ref{appA}.}
\begin{align}
\Big[2\nabla_\mu-iB_\mu-gW\gamma_\mu+4iH_{\mu\nu}\gamma^\nu\Big]\epsilon\,=&\,0\,, \notag \\
\Big[\partial_\mu{u}_i\gamma^\mu+g\partial_{u_i}W+i\partial_{u_i}H_{\mu\nu}\gamma^{\mu\nu}\Big]\epsilon\,=&\,0\,, \notag \\
\left[\partial_\mu\phi\gamma^\mu+g\partial_\phi{W}+\frac{i}{2}\partial_\phi{B}_\mu\gamma^\mu\right]\epsilon\,=&\,0\,,
\end{align}
where we define
\begin{align} \label{hhbbdef}
H_{\mu\nu}=&\,-\frac{1}{2\sqrt{2}}\left(e^{u_1+u_2+u_3}F^{-0}_{\mu\nu}+e^{u_1-u_2-u_3}F^{-1}_{\mu\nu}+e^{-u_1+u_2-u_3}F^{-2}_{\mu\nu}+e^{-u_1-u_2+u_3}F^{-3}_{\mu\nu}\right)\,, \notag \\
B_\mu\,=&\,-4\sqrt{2}gA_\mu^0+\frac{1}{2}e^{2\phi}D_\mu\sigma\,.
\end{align}

The scalar potential has a supersymmetric vacuum at, \cite{Halmagyi:2013sla},
\begin{equation} \label{radii}
e^{2u_i}\,=\,\sqrt{\frac{e_0}{6}}\,, \qquad e^{-2\phi}\,=\,\frac{e_0}{6}\,, \qquad L_{AdS_4}\,=\,\frac{1}{2}\left(\frac{e_0}{6}\right)^{3/4}\, ,
\end{equation}
whereas the scalar $\sigma$ is a flat direction and gets eaten by the massive vector $A^m$ as usual in the Higgs mechanism, \cite{Hristov:2010eu}.
The vacuum uplifts to the $AdS_4\times{Q}^{1,1,1}$ solution of eleven-dimensional supergravity and is dual to the flavored ABJM theories, \cite{Benini:2009qs, Cremonesi:2010ae}. The radius of the $AdS_4$ is given by
\begin{equation} 
L_{AdS_4}^2\,=\,-\frac{3}{g^2\mathcal{V}_*}\,,
\end{equation}
where $\mathcal{V}_*$ is the value of the scalar potential at the vacuum.~\footnote{Note that in \eqref{radii} and, as indicated, in some of the latter sections we fix $g = 1$ for simplicity.}

The consistent truncation of eleven-dimensional supergravity on $M^{1,1,1}$ manifold readily follows from the truncation on $Q^{1,1,1}$ manifold by identifying the scalar and gauge fields by
\begin{equation}
u_3\,=\,u_2\,, \qquad A^3\,=\,A^2\,.
\end{equation}
Note that there is one Betti vector in the truncation from \eqref{vectorsrmbb}, corresponding to a baryonic symmetry holographically. The $AdS_4\times{M}^{1,1,1}$ solution of eleven-dimensional supergravity is dual to 3d SCFTs studied in \cite{Martelli:2008si}.  By setting all gauge fields and corresponding scalars equal, the truncation further reduces to minimal gauged supergravity, see appendix \ref{appA}. This is also the relevant truncation for all other homogeneous SE$_7$ manifolds in \cite{Cassani:2012pj} as they only exhibit massive vectors and no additional baryonic symmetries.

From the AdS/CFT correspondence, the free energy of pure $AdS_4$ with an asymptotic boundary of $S^3$ is given by,~\footnote{We also show how to derive this formula from the gravitational block construction in section \ref{sec:6}.}
\begin{equation} \label{flads4}
\mathcal{F}_{S^3}\,=\,\frac{\pi{L}^2_{AdS_4}}{2G_N^{(4)}}\,,
\end{equation}
where $G_N^{(4)}$ is the four-dimensional Newton's constant. For the solutions of $AdS_4\times{Y}$ where $Y$ is a seven-dimensional Einstein space with $N$ units of fluxes, the holographic free energy is, \cite{Herzog:2010hf},
\begin{equation}
\mathcal{F}_{S^3}\,=\,N^{3/2}\sqrt{\frac{2\pi^6}{27\text{vol}(Y)}}\,,
\end{equation}
where the metric on $Y$ is normalized to be $R_{ij}=6g_{ij}$. Thus we obtain the holographic free energy of ABJM and 3d SCFTs dual to $AdS_4\times{Q}^{1,1,1}$ and $AdS_4\times{M}^{1,1,1}$, respectively,
\begin{align} \label{fabjm}
\mathcal{F}_{S^3}^{\text{ABJM}}\,=&\,\frac{\sqrt{2}\pi}{3}N^{3/2}\,, \qquad \qquad \qquad \,\,\, \text{vol}(S^7)\,=\,\frac{\pi^4}{3}\,, \notag \\
\mathcal{F}_{S^3}^{Q^{1,1,1}/\mathbb{Z}_n}\,=&\,\frac{4\pi}{3\sqrt{3}}n^{1/2}N^{3/2}\,, \qquad \text{vol}(Q^{1,1,1}/\mathbb{Z}_n)\,=\,\frac{\pi^4}{8n}\, \notag \\
\mathcal{F}_{S^3}^{M^{1,1,1}/\mathbb{Z}_n}\,=&\,\frac{16\pi}{9\sqrt{3}}n^{1/2}N^{3/2}\,, \qquad \text{vol}(M^{1,1,1}/\mathbb{Z}_n)\,=\,\frac{9\pi^4}{128n}\,,
\end{align}
where the free energy of flavored ABJM dual to $AdS_4\times{Q}^{1,1,1}$ was calculated in \cite{Cheon:2011vi, Jafferis:2011zi} and for 3d SCFTs dual to $AdS_4\times{M}^{1,1,1}$ in \cite{Gulotta:2011aa, Gulotta:2011si}.

\section{$AdS_2$ ansatz}
\label{sec:3}

This section, as well as the next one, follows the main outline and logic of \cite{Arav:2022lzo} and \cite{Suh:2022pkg} for a simpler comparison.

We first consider the metric and the gauge fields,
\begin{align}
ds^2\,=&\,e^{2V}ds_{AdS_2}^2+f^2dy^2+h^2dz^2\,, \notag \\
A^I\,=&\,a^Idz\,,
\end{align}
where $AdS_2$ is a unit radius metric on $AdS_2$ and $V$, $f$, $h$, and $a^I$, $I=0,\ldots,3$, as well as the scalar fields $\phi, u_i$ are functions of $y$-coordinate only. In order to avoid partial differential equations from the equations of motion for the gauge fields, the scalar field $\sigma$ is given by $\sigma=\bar{\sigma}z$ where $\bar{\sigma}$ is constant. Hence, we find
\begin{equation}
B_\mu{d}x^\mu\,\equiv\,B_zdz\,,
\end{equation}
where $B_z$ is again a function of the $y$-coordinate only.

We introduce an orthonormal frame,
\begin{equation}
e^a\,=\,e^V\bar{e}^a\,, \qquad e^2\,=\,fdy\,, \qquad e^3\,=\,hdz\,,
\end{equation}
where $\bar{e}^a$ is an orthonormal frame on $ds_{AdS_2}^2$. In the frame coordinates, the field strengths are given by
\begin{equation}
F_{23}^I\,=\,f^{-1}h^{-1}\left(a^I\right)'\,.
\end{equation}

From the gauge field equations we find the following integrals of motion,
\begin{align} \label{ef123}
-\frac{1}{\sqrt{2}}e^{2V}\left(e^{2u_1-2u_2-2u_3}F_{23}^1-e^{-2u_1+2u_2-2u_3}F_{23}^2\right)\,=\,\mathcal{E}_{F_1}\,, \notag \\
-\frac{1}{\sqrt{2}}e^{2V}\left(e^{-2u_1+2u_2-2u_3}F_{23}^2-e^{-2u_1-2u_2+2u_3}F_{23}^3\right)\,=\,\mathcal{E}_{F_2}\,, \notag \\
-\frac{1}{\sqrt{2}}e^{2V}\left(e^{-2u_1-2u_2+2u_3}F_{23}^3-e^{2u_1-2u_2-2u_3}F_{23}^1\right)\,=\,\mathcal{E}_{F_3}\,,
\end{align}
as well as
\begin{align} \label{er123}
-\frac{1}{\sqrt{2}}e^{2V}\left(e^{2u_1+2u_2+2u_3}F_{23}^0+\frac{e_0}{2}e^{2u_1-2u_2-2u_3}F_{23}^1\right)\,=\,\mathcal{E}_{R_1}\,, \notag \\
-\frac{1}{\sqrt{2}}e^{2V}\left(e^{2u_1+2u_2+2u_3}F_{23}^0+\frac{e_0}{2}e^{-2u_1+2u_2-2u_3}F_{23}^2\right)\,=\,\mathcal{E}_{R_2}\,, \notag \\
-\frac{1}{\sqrt{2}}e^{2V}\left(e^{2u_1+2u_2+2u_3}F_{23}^0+\frac{e_0}{2}e^{-2u_1-2u_2+2u_3}F_{23}^3\right)\,=\,\mathcal{E}_{R_3}\,,
\end{align}
with
\begin{equation}
\left(e^{2V-2u_1-2u_2+2u_3}F^3_{23}\right)'\,=\,\sqrt{2}ge^{2V}fh^{-1}e^{4\phi}D_z\sigma\,,
\end{equation}
where $\mathcal{E}_{F_i}$ and $\mathcal{E}_{R_i}$ are constant. Among the six integrals of motion in \eqref{ef123} and \eqref{er123}, three of them are independent.

\subsection{BPS equations}
\label{sec:3.1}

We employ the gamma matrix decomposition
\begin{equation}
\gamma^m\,=\,\Gamma^m\otimes\sigma^3\,, \qquad \gamma^2\,=\,I_2\otimes\sigma^1\,, \qquad \gamma^3\,=\,I_2\otimes\sigma^2\,,
\end{equation}
where $\Gamma^m$ are two-dimensional gamma matrices and $\sigma^i$ the Pauli matrices. The spinors are given by
\begin{equation}
\epsilon\,=\,\psi\otimes\chi\,.
\end{equation}
The two-dimensional spinor satisfies
\begin{equation}
D_m\psi\,=\,\frac{1}{2}\kappa\Gamma_m\psi\,,
\end{equation}
where $\kappa=\pm1$ fixes the chirality.

The resulting BPS equations are discussed in detail in appendix \ref{appB}, where the angular parameter $\xi$ is used to parametrize the Killing spinor projection that is required by the ansatz, see \eqref{eq:projection}-\eqref{fromxi}. For the general case of $\sin\xi\ne0$, the complete BPS equations are obtained in the appendix and are given by
\begin{align} \label{fullbps}
f^{-1}\xi'\,=&\,2gW\cos\xi+\kappa{e}^{-V}\,, \notag \\
f^{-1}V'\,=&\,gW\sin\xi\,, \notag \\
f^{-1}u_i'\,=&\,-g\partial_{u_i}W\sin\xi\,, \notag \\
f^{-1}\phi'\,=&\,-\frac{g\partial_\phi{W}}{\sin\xi}\,, \notag \\
f^{-1}\frac{h'}{h}\,=&\,\frac{1}{\sin\xi}\Big(\kappa{e}^{-V}\cos\xi+gW\left(1+\cos^2\xi\right)\Big)\,,
\end{align}
with two constraints,
\begin{align} \label{bpsconstraints}
\left(s-B_z\right)\sin\xi\,=&\,-2gWh\cos\xi-\kappa{h}e^{-V}\,, \notag \\
g\partial_\phi{W}\cos\xi\,=&\,\frac{1}{2}\partial_\phi{B}_z\sin\xi{h}^{-1}\,.
\end{align}
The field strengths of gauge fields are given by
\begin{align} \label{bpsfs}
\partial_{u_i}H_{23}\,=&\,-\frac{1}{4}g\partial_{u_i}W\cos\xi\,, \notag \\
H_{23}\,=&\,-\frac{1}{4}gW\cos\xi-\frac{1}{4}\kappa{e}^{-V}\,.
\end{align}
These BPS equations are consistent with the equations of motion from the Lagrangian in \eqref{lagmain} given in appendix \ref{appA}.

\subsection{Integrals of motion}

There is an integral of the BPS equations,
\begin{equation} \label{hevks}
he^{-V}\,=\,k\sin\xi\,,
\end{equation}
where $k$ is a constant. Thus at the poles of the spindle solutions at $h=0$, we have $\sin\xi=0$. From \eqref{fullbps} and \eqref{bpsconstraints} we find
\begin{equation} \label{xipxip}
\xi'\,=\,-k^{-1}\left(s-B_z\right)\left(e^{-V}f\right)\,,
\end{equation}
with two constraints in \eqref{bpsconstraints} to be
\begin{align} \label{constrainttwo}
\left(s-B_z\right)\,=&\,-k\Big[2gWe^V\cos\xi+\kappa\Big]\,, \notag \\
g\partial_\phi{W}\cos\xi\,=&\,\frac{1}{2}k^{-1}e^{-V}\partial_\phi{B}_z\,.
\end{align}

From the field strengths in \eqref{bpsfs}, we find expressions of the integrals of motion to be~{\footnote{Note that accidentally, when $e_0=2$, they reduce to the corresponding expressions from the mass-deformed ABJM in \cite{Suh:2022pkg}.}}
\begin{align}
\mathcal{E}_{R_1}\,=&\,e^V\left[4ge^V\cos\xi+\kappa{e}^{u_1}\cosh\left(u_2+u_3\right)\right]-\frac{2-e_0}{4}\kappa{e}^Ve^{2u_1}e^{-u_1-u_2-u_3}\,, \notag \\
\mathcal{E}_{R_2}\,=&\,e^V\left[4ge^V\cos\xi+\kappa{e}^{u_2}\cosh\left(u_3+u_1\right)\right]-\frac{2-e_0}{4}\kappa{e}^Ve^{2u_2}e^{-u_1-u_2-u_3}\,, \notag \\
\mathcal{E}_{R_3}\,=&\,e^V\left[4ge^V\cos\xi+\kappa{e}^{u_3}\cosh\left(u_1+u_2\right)\right]-\frac{2-e_0}{4}\kappa{e}^Ve^{2u_3}e^{-u_1-u_2-u_3}\,,
\end{align}
and
\begin{align}
\mathcal{E}_{F_1}\,=&\,\kappa{e}^Ve^{u_3}\sinh\left(u_1-u_2\right)\,, \notag \\
\mathcal{E}_{F_2}\,=&\,\kappa{e}^Ve^{u_1}\sinh\left(u_2-u_3\right)\,, \notag \\
\mathcal{E}_{F_3}\,=&\,\kappa{e}^Ve^{u_2}\sinh\left(u_3-u_1\right)\,.
\end{align}

\subsection{Boundary conditions for spindle solutions} \label{sec33}

We can choose the conformal gauge,
\begin{equation}
f\,=\,-e^V\,,
\end{equation}
in order to have the metric of the form,
\begin{equation}
ds^2\,=\,e^{2V}\left[ds_{AdS_2}^2+ds_{{\Sigma}}^2\right]\,,
\end{equation}
where the metric on a spindle, ${{\Sigma}}$ is
\begin{equation}
ds_{{\Sigma}}^2\,=\,dy^2+k^2\sin^2\xi{d}z^2\,.
\end{equation}
The spindle solutions have two poles at $y=y_{N,S}$ with deficit angles of $2\pi\left(1-\frac{1}{n_{N,S}}\right)$. We set the period of the azimuthal angle, $z$, to be
\begin{equation}
\Delta{z}\,=\,2\pi\,.
\end{equation}

\subsubsection{Analysis of the BPS equations}

We analyze the BPS equations for the spindle solutions. At the poles, $y\,=\,y_{N,S}$, as $k\sin\xi\rightarrow0$, we obtain $\cos\xi\rightarrow\pm1$ if $k\ne0$. Hence, we have $\cos\xi_{N,S}\,=\,(-1)^{t_{N,S}}$ with $t_{N,S}\in\{0,1\}$. We select $y_N<y_S$ and $y\in[y_N,y_S]$. We make an assumption that the deficit angles at the poles are $2\pi\left(1-\frac{1}{n_{N,S}}\right)$ with $n_{N,S}\ge1$. Then we require the metric to have $|\left(k\sin\xi\right)'|_{N,S}\,=\,\frac{1}{n_{N,S}}$. From the symmetry of BPS equations in \eqref{hsymm} and \eqref{hevks}, we further choose
\begin{equation}
h\le0\,, \qquad \Leftrightarrow \qquad k\sin\xi\le0\,.
\end{equation}
Then we obtain $\left(k\sin\xi\right)'|_N<0$ and $\left(k\sin\xi\right)'|_S>0$. Thus, we impose 
\begin{equation}
\left(k\sin\xi\right)'|_{N,S}\,=\,-\frac{(-1)^{l_{N,S}}}{n_{N,S}}\,, \qquad l_N\,=\,0\,,l_S\,=\,1\,.
\end{equation}

There are two distinct classes of spindle solutions, the twist and the anti-twist classes, \cite{Ferrero:2021etw}. The spinors are of the same chirality at the poles for the twist solutions and opposite chiralities for the anti-twist solutions,
\begin{align} \label{cosxi1}
\cos\xi|_{N,S}\,=\,(-1)^{t_{N,S}}; \qquad &\text{Twist:} \,\, \qquad \quad \left(t_N,t_S\right)\,=\,\left(1,1\right) \quad \text{or} \quad \left(0,0\right)\,, \notag \\
&\text{Anti-Twist:} \quad \left(t_N,t_S\right)\,=\,\left(1,0\right) \quad \text{or} \quad \left(0,1\right)\,.
\end{align}

As we have $\left(k\sin\xi\right)'\,=\,+\cos\xi\left(s-B_z\right)$ from the BPS equation in \eqref{xipxip}, we find
\begin{equation}
\left(s-B_z\right)|_{N,S}\,=\,\frac{1}{n_{N,S}}(-1)^{l_{N,S}+t_{N,S}+1}\,.
\end{equation}
We consider the flux quantization for R-symmetry flux. From \eqref{hhbbdef} we have $-gF^R\,=\,dB+d\left(\frac{1}{2}e^{2\phi}D\sigma\right)$. At the poles, as $\phi=0$ unless $D\sigma=0$, the second term on the right hand side of $F^R$ does not contribute to the flux quantization. Then, we find the R-symmetry flux quantized to be
\begin{equation} \label{rsymmq}
\frac{1}{2\pi}\int_{{\Sigma}}gF^\text{R}\,\equiv\,\frac{1}{2\pi}\int_{{\Sigma}}\left(-dB\right)\,=\,\frac{n_N(-1)^{t_S}+n_S(-1)^{t_N}}{n_Nn_S}\,.
\end{equation}

We have $\partial_zB=e^{2\phi}D_z\sigma$. Once again, as $\phi=0$ unless $D\sigma=0$ at the poles, we find $\partial_\phi{B}_z=0$ at the poles. From the constraint in \eqref{constrainttwo} we also find $\partial_\phi{W}=0$ at the poles. Hence, we have
\begin{equation} \label{dbdw}
\partial_\phi{B}_z|_{N,S}\,=\,\partial_\phi{W}|_{N,S}\,=\,0\,.
\end{equation}

We further assume that the hypermultiplet scalars, $(\phi,\sigma)$, are non-vanishing at the poles and we find
\begin{equation} \label{chitopsi}
\phi|_N\,,\phi|_S\,\ne\,0\,, \qquad \Rightarrow \qquad D_z\sigma|_N\,=\,D_z\sigma|_S\,=\,0\,.
\end{equation}
Thus, we find that the flux charging $\sigma$ should vanish,
\begin{equation} \label{msymmq}
\frac{1}{2\pi}\int_{{\Sigma}}\sqrt{2}g\left(e_0F^0-2F^1-2F^2-2F^3\right)\,=\,\left(D_z\sigma\right)|_{y_N}^{y_S}\,=\,0\,.
\end{equation}

From \eqref{chitopsi} and the second equation in \eqref{dbdw} we obtain
\begin{equation} \label{dw=0}
\left.\left(e^{-2u_1}+e^{-2u_2}+e^{-2u_3}-\frac{e_0}{2}e^{-2u_1-2u_2-2u_3}\right)\right|_{N,S}\,=\,0\,, \quad \Rightarrow \quad W|_{N,S}\,=\,2e^{-u_1-u_2-u_3}|_{N,S}\,.
\end{equation}
We introduce quantities,
\begin{equation} \label{defm1m2}
M_{(1)}\,\equiv\,-2ge^Ve^{-u_1-u_2-u_3}\,, \qquad M_{(2)}\,\equiv\,-\frac{\kappa}{4}M_{(1)}+M_{(1)}^2\cos\xi\,,
\end{equation}
where $M_{(1)}<0$. We eliminate $V$ by the first equation in \eqref{constrainttwo} and eliminate $\cos\xi$ by \eqref{cosxi1}. Then the integrals of motion are given by
\begin{equation}
\mathcal{E}_{R_i}\,=\,-\frac{e_0}{2}\frac{\kappa}{4g}M_{(1)}e^{2u_i}+\frac{1}{g}M_{(2)}e^{2u_1+2u_2+2u_3}\,,
\end{equation}
where we have
\begin{align}
M_{(1)}\,=&\,\frac{1}{2}(-1)^{t_{N,S}}\kappa-\frac{1}{2kn_{N,S}}(-1)^{l_{N,S}}\,, \notag \\
M_{(2)}\,=&\,\frac{1}{8}(-1)^{t_{N,S}}+\frac{1}{4k^2n_{N,S}^2}(-1)^{t_{N,S}}-\frac{3\kappa}{8kn_{N,S}}(-1)^{l_{N,S}}\,.
\end{align}
Lastly, we eliminate one of the scalar fields, $u_i$, say $u_3$, by the condition on the left hand side of \eqref{dw=0}. Thus we have three independent integrals of motion in terms of two scalar fields, $u_1$ and $u_2$. As the integrals of motion are constant and have identical values at the poles, we find three algebraic constraints with four unknowns, $\left(u_{1N},u_{1S},u_{2N},u_{2S}\right)$, 
\begin{align} \label{assoeq}
\mathcal{E}_{R_1}\left(u_{1N},u_{2N}\right)\,=&\,\mathcal{E}_{R_1}\left(u_{1S},u_{2S}\right)\,, \notag \\
\mathcal{E}_{R_2}\left(u_{1N},u_{2N}\right)\,=&\,\mathcal{E}_{R_2}\left(u_{1S},u_{2S}\right)\,, \notag \\
\mathcal{E}_{R_3}\left(u_{1N},u_{2N}\right)\,=&\,\mathcal{E}_{R_3}\left(u_{1S},u_{2S}\right)\,.
\end{align}
In the next subsection, we find additional constraints to fix all the values of the fields at the poles.

\subsubsection{Fluxes}

In appendix \ref{appB}, the field strengths are expressed in terms of the scalar fields, metric functions, the angle, $\xi$, and constant, $k$,
\begin{equation}
F^I_{yz}\,=\,\left(a^I\right)'\,=\,\left(\mathcal{I}^{(I)}\right)'\,,
\end{equation}
where we define
\begin{align} \label{idef}
\mathcal{I}^{(0)}\,\equiv&\,\frac{1}{\sqrt{2}}gke^V\cos\xi{e}^{-u_1-u_2-u_3}\,, \notag \\
\mathcal{I}^{(1)}\,\equiv&\,\frac{1}{\sqrt{2}}gke^V\cos\xi{e}^{-u_1+u_2+u_3}\,, \notag \\
\mathcal{I}^{(2)}\,\equiv&\,\frac{1}{\sqrt{2}}gke^V\cos\xi{e}^{u_1-u_2+u_3}\,, \notag \\
\mathcal{I}^{(3)}\,\equiv&\,\frac{1}{\sqrt{2}}gke^V\cos\xi{e}^{u_1+u_2-u_3}\,.
\end{align}
The fluxes are solely determined by the data at the poles,
\begin{equation}
\frac{p_I}{n_Nn_S}\,\equiv\,\frac{1}{2\pi}\int_{\Sigma}gF^I\,=\,g\mathcal{I}_I|_N^S\,.
\end{equation}

From \eqref{vectorsrmbb} we define R-symmetry, massive vector and two Betti vector fluxes, respectively,
\begin{align}
\mathcal{I}_\text{R}|_{N,S}\,\equiv&\,4\sqrt{2}\mathcal{I}^{(0)}|_{N,S}\,, \notag \\
\mathcal{I}_m|_{N,S}\,\equiv&\,\sqrt{2}\left(e_0\mathcal{I}^{(0)}-2\mathcal{I}^{(1)}-2\mathcal{I}^{(2)}-2\mathcal{I}^{(3)}\right)|_{N,S}\,, \notag \\
\mathcal{I}_{\mathcal{B}_1}|_{N,S}\,\equiv&\,4\sqrt{2}\left(\mathcal{I}^{(1)}-2\mathcal{I}^{(2)}+\mathcal{I}^{(3)}\right)|_{N,S}\,, \notag \\
\mathcal{I}_{\mathcal{B}_2}|_{N,S}\,\equiv&\,4\sqrt{2}\left(\mathcal{I}^{(1)}-2\mathcal{I}^{(3)}\right)|_{N,S}\,.
\end{align}
Employing \eqref{idef} and \eqref{dw=0}, we obtain
\begin{align}
g\mathcal{I}_\text{R}|_{N,S}\,=&\,4\sqrt{2}g\mathcal{I}^{(0)}|_{N,S} \notag \\
=&\,4gke^V\cos\xi{e}^{-u_1-u_2-u_3}|_{N,S} \notag \\
=&\,-2kM_{(1)}(-1)^{t_{N,S}+1}|_{N,S}\,, \\
\notag \\
g\mathcal{I}_m|_{N,S}\,=&\,\sqrt{2}g\left(e_0\mathcal{I}^{(0)}-2\mathcal{I}^{(1)}-2\mathcal{I}^{(2)}-2\mathcal{I}^{(3)}\right)|_{N,S} \notag \\
=&\,gke^V\cos\xi\left(e_0e^{-u_1-u_2-u_3}-2e^{-u_1+u_2+u_3}-2e^{u_1-u_2+u_3}-2e^{u_1+u_2-u_3}\right)|_{N,S} \notag \\
=&\,gke^V\cos\xi\,e^{u_1+u_2+u_3}\left(e_0e^{-2u_1-2u_2-2u_3}-2e^{-2u_1}-2e^{-2u_2}-2e^{-2u_3}\right)|_{N,S}\,=\,0\,,
\end{align}
where we employed \eqref{dw=0}. Then we recover the R-symmetry flux quantization, \eqref{rsymmq}, and the vanishing of the flux of massive vector field, \eqref{msymmq}, respectively,
\begin{align}
g\mathcal{I}_\text{R}|_N^S\,=&\,\frac{n_N(-1)^{t_S}+n_S(-1)^{t_N}}{n_Nn_S}\,, \notag \\
g\mathcal{I}_m|_N^S\,=&\,0\,.
\end{align}
The fluxes of two Betti vector fields are 
\begin{align} \label{fsymmq1}
\frac{p_{\mathcal{B}_1}}{n_Nn_S}\,\equiv&\,g\mathcal{I}_{\mathcal{B}_1}|_N^S\,=\,4\sqrt{2}g\left(\mathcal{I}^{(1)}-2\mathcal{I}^{(2)}+\mathcal{I}^{(3)}\right)|_N^S \notag \\
=&\,4gke^V\cos\xi\left(e^{-u_1+u_2+u_3}-2e^{u_1-u_2+u_3}+e^{u_1+u_2-u_3}\right)|_N^S \notag \\
=&\,2kM_{(1)}(-1)^{t_{N,S}}\left(e^{2u_2+2u_3}-2e^{2u_1+2u_3}+e^{2u_1+2u_2}\right)|_N^S\,,\\
\notag \\ \label{fsymmq2}
\frac{p_{\mathcal{B}_2}}{n_Nn_S}\,\equiv&\,g\mathcal{I}_{\mathcal{B}_2}|_N^S\,=\,4\sqrt{2}g\left(\mathcal{I}^{(1)}-\mathcal{I}^{(3)}\right)|_N^S \notag \\
=&\,4gke^V\cos\xi\left(e^{-u_1+u_2+u_3}-e^{u_1+u_2-u_3}\right)|_N^S \notag \\
=&\,2kM_{(1)}(-1)^{t_{N,S}}\left(e^{2u_2+2u_3}-e^{2u_1+2u_2}\right)|_N^S\,,
\end{align}
where $p_{\mathcal{B}_1}$ and $p_{\mathcal{B}_2}$ are integers. The expressions of $k$ and one of the scalar fields, $u_i$, which was not fixed in \eqref{assoeq} is determined by these two constraints.

{\bf Summary of the constraints to determine all the boundary conditions:} We summarize the constraints obtained to determine all the boundary conditions. By solving seven associated equations, the left hand side of \eqref{dw=0}, \eqref{assoeq}, \eqref{fsymmq1}, and \eqref{fsymmq2}, we can determine the values of the scalar fields, $u_1$, $u_2$, $u_3$, at the north and south poles and also the constant, $k$, in terms of $n_{N,S}$, $t_{N,S}$, $p_{\mathcal{B}_1}$, and $p_{\mathcal{B}_2}$. Then the values of the metric function, $V$, at the poles are determined from the definition of $M_{(1)}$ in \eqref{defm1m2}. This fixes all the boundary conditions except the hyper scalar field, $\phi$, which will be chosen when constructing the solutions explicitly. However, the constraint equations are quite complicated and it appears to be not easy to solve them.

Even though we are not able to solve for the boundary conditions in terms of $n_{N,S}$, $t_{N,S}$, $p_{\mathcal{B}_1}$, and $p_{\mathcal{B}_2}$ analytically, if we choose numerical values of $n_{N,S}$, $t_{N,S}$, $p_{\mathcal{B}_1}$, and $p_{\mathcal{B}_2}$, the constraints can be solved to determine all the boundary conditions. For instance, in the anti-twist class, for the choice of
\begin{align} \label{ninput}
n_N\,=&\,4\,, \qquad n_S\,=\,1\,, \qquad p_{\mathcal{B}_1}\,=\,1\,, \qquad p_{\mathcal{B}_2}\,=\,2\,, \notag \\
g\,=&\,1\,, \qquad \kappa\,=\,+1\,, \qquad e_0\,=\,2\,,
\end{align}
we find the boundary conditions to be
\begin{align} \label{noutput}
e^{2u_{1N}}\,\approx&\,0.174581\,, \qquad \,\,\,\,\,\, e^{2u_{1S}}\,\approx\,0.0768503\,, \notag\\
e^{2u_{2N}}\,\approx&\,0.418164\,, \qquad \quad \,\,\, e^{2u_{2S}}\,\approx\,0.411234\,, \notag\\
e^{2u_{3N}}\,\approx&\,1.5639\,, \qquad \qquad e^{2u_{3S}}\,\approx\,1.98408\,, \notag\\
k\,\approx&\,-0.683914\,.
\end{align}
In this way, without finding analytic expression of the Bekenstein-Hawking entropy, we can determine numerical value for each choice of $n_{N,S}$, $t_{N,S}$, $p_{\mathcal{B}_1}$, and $p_{\mathcal{B}_2}$. Furthermore, we will be able to construct the solutions explicitly numerically, see below and in the attached {\it Mathematica} file.

\subsubsection{The Bekenstein-Hawking entropy} \label{sec333}

The $AdS_2\times{\Sigma}$ solution would be the horizon of a presumed black hole which asymptotes to the $AdS_4\times{Q}^{1,1,1}$ vacuum of M-theory. We calculate the Bekenstein-Hawking entropy of the presumed black hole solution.

From the AdS/CFT dictionary, \eqref{flads4} and \eqref{fabjm}, the four-dimensional Newton's constant is
\begin{equation}
\frac{1}{2G_N^{(4)}}\,=\,\left(\frac{6}{e_0}\right)^{3/2}\frac{16}{3\sqrt{3}}N^{3/2}\,.
\end{equation}
Then the two-dimensional Newton's constant is given by
\begin{equation}
\left(G_N^{(2)}\right)^{-1}\,=\,\left(G_N^{(4)}\right)^{-1}\Delta{z}\int_{y_N}^{y_S}|fh|dy\,.
\end{equation}
Employing the BPS equations, we find
\begin{equation}
fh\,=\,ke^Vf\sin\xi\,=\,+\frac{k}{\kappa}\left(e^{2V}\cos\xi\right)'\,.
\end{equation}
Thus the Bekenstein-hawking entropy is solely expressed by the data at the poles,
\begin{align} \label{bhent}
&S_{\text{BH}}\,=\,\frac{1}{4G_N^{(2)}}\,=\,\left(\frac{6}{e_0}\right)^{3/2}\frac{16\pi}{3\sqrt{3}}N^{3/2}g^2\left(-\frac{k}{\kappa}\right)\left[e^{2V}\cos\xi\right]_N^S \notag \\
&=\,-\left(\frac{6}{e_0}\right)^{3/2}\frac{16N^{3/2}k}{3\sqrt{3}\kappa}\left(\frac{1}{4}M_{(1)}^2|_S\,e^{2u_{1S}+2u_{2S}+2u_{3S}}(-1)^{t_S}-\frac{1}{4}M_{(1)}^2|_N\,e^{2u_{1N}+2u_{2N}+2u_{3N}}(-1)^{t_N}\right)\,,
\end{align}
where we expressed the Bekenstein-Hawking entropy in terms of $M_{(1)}$.

As we can determine the numerical values of the boundary conditions for each choice of $n_{N,S}$, $t_{N,S}$, $p_{\mathcal{B}_1}$, and $p_{\mathcal{B}_2}$, we can find the numerical value of the Bekenstein-Hawking entropy as well. For instance, for the choice of \eqref{ninput}, the Bekenstein-Hawking entropy is given by $S_{\text{BH}}\,\approx\,0.127521N^{3/2}$.

Furthermore, when there is no flavor charges, $p_{\mathcal{B}_1}=p_{\mathcal{B}_2}=0$, we perform a non-trivial check that the numerical value of the Bekenstein-Hawking entropy precisely matches the value obtained from the formula given in \eqref{minent} for the solutions from minimal gauged supergravity.

\section{Solving the BPS equations} \label{sec:4}

\subsection{Analytic solutions for minimal gauged supergravity} \label{sec41}

In minimal gauged supergravity associated with the $AdS_4\times{Q}^{1,1,1}$ vacuum, utilizing the class of $AdS_2\times{\Sigma}$ solutions in \cite{Ferrero:2020twa}, we find solutions in the anti-twist class to the BPS equations in \eqref{fullbps}, \eqref{bpsconstraints} and \eqref{bpsfs}.
We set $e_0=6$ as in appendix \ref{appA1}. The scalar fields take the value at the $AdS_4\times{Q}^{1,1,1}$ vacuum,
\begin{equation}
e^{2u_i}\,=\,\sqrt{\frac{e_0}{6}}\,, \qquad e^{-2\phi}\,=\,\frac{e_0}{6}\,, \qquad L_{AdS_4}\,=\,\frac{1}{2}\left(\frac{e_0}{6}\right)^{3/4}\,.
\end{equation}
The metric and the gauge field are given by
\begin{align}
ds^2\,=&\,\frac{1}{2g^2}\left[\frac{y^2}{4}ds_{AdS_2}^2+\frac{y^2}{q(y)}dy^2+\frac{q(y)}{4y^2}c_0^2dz^2\right]\,, \notag \\
A^0\,=\,A^1\,=\,A^2\,=\,A^3\,=&\,-\left[\frac{c_0\kappa}{2g}\left(1-\frac{a}{y}\right)+\frac{s}{g}\right]dz\,,
\end{align}
and we have
\begin{equation}
\sin\xi\,=\,-\frac{\sqrt{q(y)}}{y^2}\,, \qquad \cos\xi\,=\,\kappa\frac{2y-a}{y^2}\,.
\end{equation}
Note that for the overall factor in the metric, we have $L_{AdS_4}^2=\frac{1}{2g^2}$ for the $AdS_4\times{Q}^{1,1,1}$ vacuum from \eqref{radii}. The quartic function is given by
\begin{equation}
q(y)\,=\,y^4-4y^2+4ay-a^2\,,
\end{equation}
and the constants are
\begin{align}
a\,=&\,\frac{n_S^2-n_N^2}{n_S^2+n_N^2}\,, \notag \\
c_0\,=&\,\frac{\sqrt{n_S^2+n_N^2}}{\sqrt{2}n_Sn_N}\,.
\end{align}
We set $n_S>n_N$. For the two middle roots of $q(y)$, $y\in[y_N,y_S]$, we find
\begin{equation}
y_N\,=\,-1+\sqrt{1+a}\,, \qquad y_S\,=\,1-\sqrt{1-a}\,.
\end{equation}
The Bekenstein-Hawking entropy is calculated to give
\begin{align} \label{minent}
S_{\text{BH}}\,=&\,\frac{\sqrt{2}\sqrt{n_S^2+n_N^2}-\left(n_S+n_N\right)}{n_Sn_N}\frac{\pi{L}_{AdS_4}^2}{4G_N^{(4)}} \notag \\
=&\,\frac{\sqrt{2}\sqrt{n_S^2+n_N^2}-\left(n_S+n_N\right)}{n_Sn_N}\frac{1}{2}\mathcal{F}_{S^3}^{Q^{1,1,1}}\,,
\end{align}
where we employed \eqref{radii} and \eqref{flads4} and $\mathcal{F}_{S^3}^{Q^{1,1,1}}=\frac{4\pi}{3\sqrt{3}}N^{3/2}$, \eqref{fabjm}, is the free energy of flavored ABJM theory.

\subsection{Numerical solutions for $p_{\mathcal{B}_1},p_{\mathcal{B}_2}\ne0$}

In section \ref{sec33}, although we were not able to find the analytic expressions of the boundary conditions, we were able to determine the numerical values of the boundary conditions for each choice of $n_{N,S}$, $t_{N,S}$, $p_{\mathcal{B}_1}$, and $p_{\mathcal{B}_2}$. Employing these results for the boundary conditions, we can numerically construct $AdS_2\times{\Sigma}$ solutions in the anti-twist class by solving the BPS equations.~{\footnote{As we do not know the analytic expressions of the boundary conditions, we could not exclude the existence of solutions in the twist class. However, we were not able to find any boundary conditions for numerical solutions in the twist class. In fact, we found solutions with the Bekenstein-Hawking entropy matching the result of gravitational block calculations. However, the scalar fields of the solutions were not real.}}

In order to solve the BPS equations numerically, we start the integration at $y=y_N$ and we choose $y_N=0$. At the poles we have $\sin\xi=0$. We scan over the initial value of $\phi$ at $y=y_N$ in search of a solution for which we have $\sin\xi=0$ in a finite range, $i.e.$, at $y=y_S$. If we find compact spindle solution, our boundary conditions guarantee the fluxes to be properly quantized.

\begin{figure}[t]
\begin{center}
\includegraphics[width=2.8in]{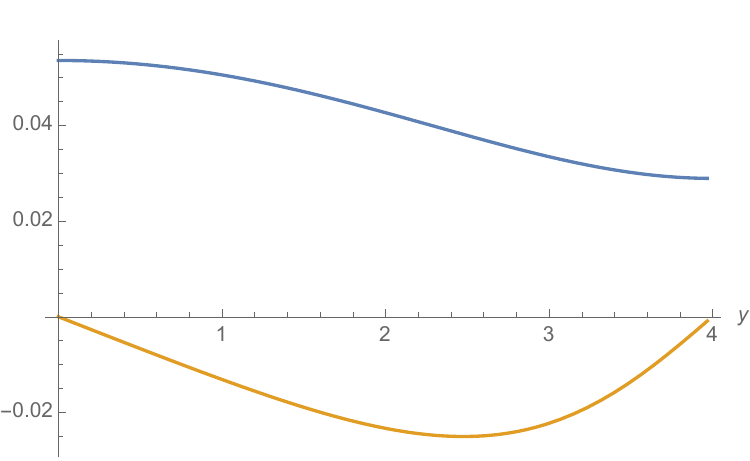} \qquad \includegraphics[width=2.8in]{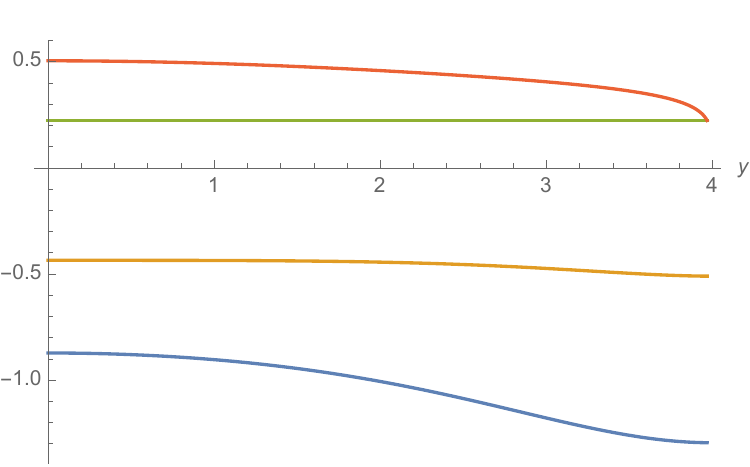}
\caption{{\it A representative $AdS_2\times{\Sigma}\times{Q^{1,1,1}}$ solution in the anti-twist class for $n_N=4$, $n_S=1$, $p_{\mathcal{B}_1}=1$ and $p_{\mathcal{B}_2}=2$ in the range of $y=[y_N,y_S]\approx[0,3.96758]$. The metric functions, $e^V$ (Blue) and $h$ (Orange), are on the left. The scalar fields, $u_1$ (Blue), $u_2$ (Orange), $u_3$ (Green), and $\phi$ (Red) are on the right. Note that $h$ vanishes at the poles.}\label{fig1}}
\end{center}
\end{figure}

We numerically perform the Bekenstein-Hawking entropy integral in section \ref{sec333} and the result matches the Bekenstein-Hawking entropy in \eqref{bhent} with the numerical accuracy of order $10^{-3}$. We present a representative solution in figure \ref{fig1} for the choice in \eqref{ninput} in the range of $y=[y_N,y_S]\approx[0,3.96758]$. The scalar field, $\phi$, takes the values, $\phi|_N\approx0.5025$ and $\phi|_S\approx0.223216$, at the poles. Note that $h$ vanishes at the poles.

There appears to be constraints on the parameter space of $n_{N,S}$, $t_{N,S}$, $p_{\mathcal{B}_1}$, and $p_{\mathcal{B}_2}$. However, without the analytic expressions of the boundary conditions, it is not easy to specify the constraints.

\section{Spindle black holes in $AdS_4\times{M}^{1,1,1}$}
\label{sec:5}

So far we have constructed the spindle black hole solutions in $AdS_4\times{Q}^{1,1,1}$. In this section, we consider the spindle black holes in $AdS_4\times{M}^{1,1,1}$.

The consistent truncation of eleven-dimensional supergravity on $M^{1,1,1}$ manifold is obtained from the truncation on $Q^{1,1,1}$ manifold by identifying the scalar and gauge fields by
\begin{equation}
u_3\,=\,u_2\,, \qquad A^3\,=\,A^2\,,
\end{equation}
and all the action, equations of motion, BPS equations, and constraints for the boundary conditions follow from this accordingly. Note that there is one Betti vector in the truncation from \eqref{vectorsrmbb}.

\begin{figure}[t]
\begin{center} 
\includegraphics[width=2.8in]{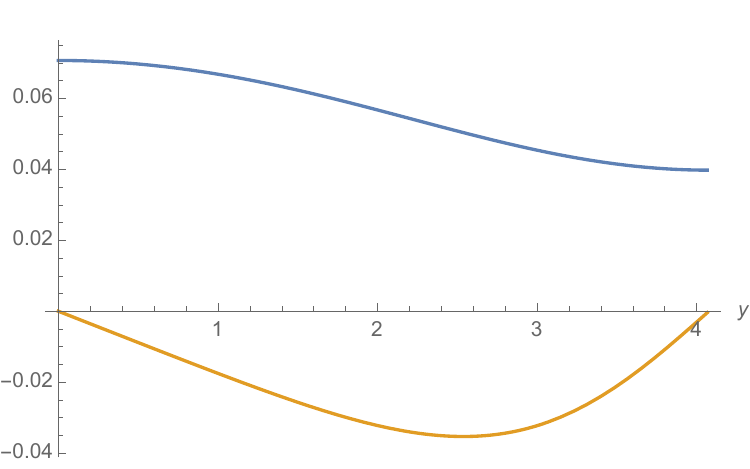} \qquad \includegraphics[width=2.8in]{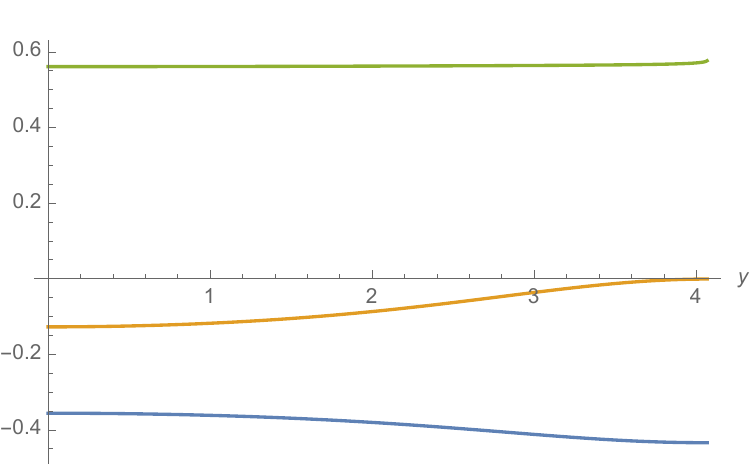}
\caption{{\it A representative $AdS_2\times{\Sigma}\times{M}^{1,1,1}$ solution in the anti-twist class for $n_N=4$, $n_S=1$, $p_{\mathcal{B}_1}=1$ in the range of $y=[y_N,y_S]\approx[0,4.06741]$. The metric functions, $e^V$ (Blue) and $h$ (Orange), are on the left. The scalar fields, $u_1$ (Blue), $u_2$ (Orange), and $\phi$ (Green) are on the right. Note that $h$ vanishes at the poles.}\label{fig2}}
\end{center}
\end{figure}

We present a representative solution numerically. For the choice of the parameters,
\begin{align} \label{m111choice}
n_N\,=&\,4\,, \qquad n_S\,=\,1\,, \qquad p_{\mathcal{B}_1}\,=\,1\,, \notag \\
g\,=&\,1\,, \qquad \kappa\,=\,+1\,, \qquad e_0\,=\,2\,,
\end{align}
we find the boundary conditions to be
\begin{align} 
e^{2u_{1N}}\,\approx&\,0.49043\,, \qquad \,\,\,\,\,\, e^{2u_{1S}}\,\approx\,0.436592\,, \notag\\
e^{2u_{2N}}\,\approx&\,0.774299\,, \qquad \quad \,\,\, e^{2u_{2S}}\,\approx\,0.926937\,, \notag\\
k\,\approx&\,-0.725009\,.
\end{align}
We present numerical solutions in figure \ref{fig2}. The scalar field, $\phi$, takes the values, $\phi|_N\approx0.561$ and $\phi|_S\approx0.575978$, at the poles. Due to \eqref{flads4} and \eqref{fabjm}, the formula for Bekenstein-Hawking entropy in \eqref{bhent} should be multiplied by $4/3$. Then the Bekenstein-Hawking entropy is calculated from \eqref{bhent} and also from numerically integrating the numerical solutions and they match precisely with the numerical accuracy of order $10^{-3}$. For the choice of \eqref{m111choice}, the Bekenstein-Hawking entropy is given by $S_\text{BH}\approx0.319973N^{3/2}$.

\section{Gravitational blocks}
\label{sec:6}

In this section, we briefly introduce the entropy functions from gluing gravitational blocks and show that extremization of entropy function correctly reproduces the Bekenstein-Hawking entropy and the scalar values at the poles of the spindle black holes we constructed.  

The main logic was already outlined in the introduction, see \eqref{eq:1}-\eqref{eq:3}. Here we are going to extend the discussion by starting with a single gravitational block, defined in \cite{Hosseini:2019iad},~\footnote{Here we use slightly different conventions and notations for the fugacities $\varphi^I, \epsilon$.} 
\be
	\cB(X^I, \epsilon) = \frac{i\, \pi}{8\, G_N^{(4)}}\, \frac{F (X^I)}{\epsilon}\ ,
\ee
where $F$ is the prepotential defining the vector multiplet scalar manifold, \cite{Hosseini:2019iad}, and $\epsilon$ can be understood geometrically as an $\Omega$-deformation parameter at each fixed point of the {\it canonical} isometry on a given background, \cite{BenettiGenolini:2019jdz,Hristov:2021qsw}. The on-shell action of a 4d BPS background $M_4$ with positive Euler number $\chi_E (M_4) > 0$ corresponding to the number of fixed points, is then given by the general {\it gluing} formula
\be
	\cF_{\partial M_4} (\varphi, \epsilon) = \sum_{s = 1}^{\chi_E (M_4)}\, \sigma_{(s)}\, \cB(X^I_{(s)}, \epsilon_{(s)})\ ,
\ee
where the corresponding identifications of $\epsilon, X^I (\varphi^I, \epsilon)$, and the relative sign $\sigma$ at each different fixed point $s$, together with one overall constraint on the fugacities coming from supersymmetry, are known as a {\it gluing rule} and depend on the particular background. 

Apart from the spindle rules that we come to shortly, for illustrative and notational purposes let us describe the simplest case of Euclidean AdS$_4$ with round S$^3$ boundary exhibiting a single fixed point. The corresponding on-shell action $\cF_{S^3}$ is precisely equal to the free energy of the dual theories on the round three-sphere discussed  in section \ref{sec:2}. Using the gluing rule in \cite{Hristov:2021qsw},~\footnote{This rule was proven to hold {\it off-shell} in the presence of vector multiplets and arbitrary higher-derivative terms in \cite{Hristov:2022plc} including supersymmetry-preserving squashing of the sphere.} we find
\be
\label{eq:S3freenergy}
	\cF_{S^3} (\varphi) = \cB(X^I =2 \varphi^I, \epsilon = 1) =  \frac{i\, \pi}{2 G_N^{(4)}}\, F(\varphi^I)\ ,
\ee
under the constraint $\xi_I \varphi^I = 2$, where in the latter equality we used the homogeneity of the prepotential.~\footnote{As discussed in the introduction and implemented below, each massive vector will add more constraints.} The extremization of the above functional with respect to $\varphi^I$ is the supergravity equivalent of the so-called F-extremization, \cite{Jafferis:2010un,Jafferis:2011zi}, and the on-shell value matches with \eqref{flads4} as we show below for the models of interest.

Unlike the simple example above, black holes with spindle horizons correspond to $\chi_E = 2$ and therefore have two fixed points that are situated precisely at the horizon that we have discussed, at the centre of the AdS$_2$ factor and the two conical singularities (or poles) of $\Sigma$. The gluing rules in this case feature the topological numbers of the spindle, as well as the magnetic fluxes of the particular background, as discussed in \cite{Hosseini:2021fge,Faedo:2021nub}. We have implemented this procedure to arrive at formulae \eqref{eq:1}-\eqref{eq:3}, where we used the notation $I^\sigma := \cF^\sigma_{S^1 \times \Sigma}$, the relative sign between the two blocks reflecting the way supersymmetry is preserved, \cite{Ferrero:2021etw}.
 In absence of electric charges and angular momentum, the Bekenstein-Hawking entropy is then obtained by extremizing the off-shell entropy function,
\begin{equation}
\label{eq:entropyextrem}
S^\sigma \left(\varphi,\epsilon;\mathfrak{n}_+,\mathfrak{n}_-\right)\,=\,  -\frac{i\, \pi}{8\, G^{(4)}_N\, \epsilon}\, \Big( F \left(\varphi^I +\epsilon\, \frak{n}^I \right)-\sigma\, F \left(  \varphi^I -\epsilon\, \frak{n}^I \right)\Big)\,.
\end{equation}
Notice again that we have $\sigma=+1$ and $\sigma=-1$ for twist and anti-twist solutions, respectively.~{\footnote{In this section we deal with $\sigma=\pm1$. This should be not confused with the hypermultiplet scalar field, $\sigma$, in the previous sections.}} The BPS condition on the $U(1)$ R-symmetry magnetic flux through the spindle is given by
\begin{equation}\label{blocktwist}
\frak{n}^R = \frac{1}{2\pi}\int_{\Sigma}dA^R\,=\,\frac{n_++\sigma\, {n}_-}{n_+n_-}\,,
\end{equation}
where $n_+$ and $n_-$ are the orbifold numbers of spindle. The variables $\varphi^I$ via the corresponding R-symmetry direction satisfy the corresponding BPS constraint,
\begin{equation} \label{varphicst}
\varphi^R-\frac{n_+-\sigma{n}_-}{n_+n_-}\epsilon\,=\,2\, .
\end{equation}
The analogous BPS constraints for the massive vector fluxes $\frak{n}^m$ and corresponding variables $\varphi^m$ instead set both of them to zero. 
We move to implement explicitly the constrained extremizations of \eqref{eq:S3freenergy} and \eqref{eq:entropyextrem} for the models of interest, starting from minimal supergravity and the STU model dual to ABJM theory and moving to the novel cases with massive vectors of mABJM and $Q^{1,1,1}$ and $M^{1,1,1}$.

\subsection{The STU model and minimal gauged supergravity}
\label{sec:61}
Let us illustrate how the gravitational block procedure works in the simplest possible case. We take the example of the pure gauged STU model (without any hypermultiplets and therefore no massive vectors) dual to ABJM theory, the prepotential is given by
\begin{equation}
\label{eq:prepabjm}
F = - 2 i\, \sqrt{X^0 X^1 X^2 X^3}\ ,
\end{equation}
and the constant gauging parameters~\footnote{Everywhere in this section we choose $g=1$ for simplicity.} $\xi_0= \xi_1 =\xi_2 = \xi_3 = 1$ define the R-symmetry vector field
\be
	A^R = A^0 + A^1 + A^2+A^3\ .
\ee
This choice of normalization for the Fayet-Iliopoulos parameters $\xi_I$ sets the AdS$_4$ scale of this model to $L_{AdS_4} = 1/\sqrt{2}$. 

Let us first evaluate the three-sphere free energy in this model, \eqref{eq:S3freenergy}, which gives
\be
	\cF^\text{ABJM}_{S^3} (\varphi) = \frac{\pi}{G_N^{(4)}}\, \sqrt{\varphi^0 \varphi^1 \varphi^2 \varphi^3}\ ,
\ee 
under the constraint $\sum_{I=0}^3 \varphi^I = 2$. It is easy to extremize the above formula, finding simply $\bar \varphi^I = 1/2$ for all $I$, which corresponds to the superconformal point (with no massive deformations) of ABJM theory. We therefore find 
\begin{equation} \label{b3abjm}
\cF^\text{ABJM}_{S^3} := \cF^\text{ABJM}_{S^3} (\bar \varphi) = \frac{\pi}{4\, G^{(4)}_N} = \frac{\pi\, L^2_{AdS_4}}{2\, G^{(4)}_N}\,,
\end{equation}
as anticipated in \eqref{flads4}. This in turn relates to the gauge group rank of ABJM via \eqref{fabjm}.

Writing the full spindle entropy function for the ABJM model can be done straightforwardly starting from \eqref{eq:entropyextrem}, but performing the actual extremization does not seem feasible analytically in the general case. It is also out of the scope of the present work to perform the match with the known explicit solutions, and therefore we just limit ourselves to the case of minimal gauged supergravity setting to zero all flavour charges. For the solutions of minimal gauge supergravity, $\mathfrak{n}\equiv\mathfrak{n}^0=\mathfrak{n}^1=\mathfrak{n}^2=\mathfrak{n}^3$, the twist condition, \eqref{blocktwist}, gives
\begin{equation}
\mathfrak{n}\,=\,\frac{n_++\sigma{n}_-}{4n_+n_-}\,,
\end{equation}
and, for $\varphi\equiv\varphi^0=\varphi^1=\varphi^2=\varphi^3$, we find from \eqref{varphicst},
\begin{equation}
\varphi-\frac{n_+-\sigma{n}_-}{4n_+n_-}\epsilon\,=\,\frac{1}{2}\,.
\end{equation}
We choose $\sigma = -1$ for the entropy function,
\begin{equation}
S^- \left(\epsilon; n_+, n_-\right)\,=\,-\frac{\pi}{4\, G_N^{(4)}\, \epsilon}\Big(\left(\varphi+\epsilon\mathfrak{n}\right)^2+\left(\varphi-\epsilon\mathfrak{n}\right)^2\Big)\,=\,-\frac{\pi}{2\, G_N^{(4)}\, \epsilon}\left(\varphi^2+\epsilon^2\mathfrak{n}^2\right)\,,
\end{equation}
with $\varphi$ and $\frak{n}$ given above in terms of $\epsilon$ and $n_\pm$. The entropy function is extremized at the value of $\epsilon$,
\begin{equation}
\bar \epsilon\,=\,\pm\frac{\sqrt{2}n_+n_-}{\sqrt{n_+^2+n_-^2}}\,,
\end{equation}
and only the lower sign gives positive entropy. We thus recover positive value of the Bekenstein-Hawking entropy of the spindle solution in minimal supergravity,
\begin{equation}
\label{eq:universalspindle}
S^-\left(\bar \epsilon;n_+, n_-\right)\,=\, \frac{\sqrt{2}\sqrt{n_+^2+n_-^2}-\left(n_++n_-\right)}{2n_+n_-}\, \mathcal{F}_{S^3}^\text{ABJM}\,,
\end{equation}
as confirmed by direct evaluation of the entropy, \cite{Ferrero:2020twa}.

Comparing the explicit form of \eqref{eq:S3freenergy} and \eqref{eq:entropyextrem}, one can see that the minimal gauged supergravity limit of every model will satisfy the same universal relation as \eqref{eq:universalspindle}, with the corresponding $\cF_{S^3}$, as also confirmed by the explicit solution in section \ref{sec41}.

\subsection{Mass-deformed ABJM}

The spindle black hole solutions from mass-deformed ABJM were obtained in \cite{Suh:2022pkg}.~\footnote{See also \cite{Bobev:2018uxk} for black holes with the horizons of $AdS_2\times\Sigma_\mathfrak{g}$ in this model where $\Sigma_\mathfrak{g}$ is a Riemann surface with genus, $\mathfrak{g}$.} The model is the same as ABJM as described right above, apart from the presence of hypermultiplet gauging and in turn one massive vector. The prepotential is therefore given by
\begin{equation}
F = - 2 i\, \sqrt{X^0 X^1 X^2 X^3}\ ,
\end{equation} 
and the R-symmetry and massive vector are given, respectively, by 
\begin{equation}
A^R = A^0 + A^1 + A^2+A^3\ , \qquad A^m = A^0 - A^1 - A^2 - A^3\ ,
\end{equation}
which corresponds to $\xi_0= \xi_1 =\xi_2 = \xi_3 = 1$ and $\zeta_0 = -\zeta_1 = -\zeta_2=-\zeta_3 = 1$ in the notation of the introductory section. This choice sets the AdS$_4$ scale of the model to $L^2_{AdS_4} = 2/3 \sqrt{2}$, see \cite{Suh:2022pkg}. 

We can directly apply the massive vector constraint on the level of the prepotential, setting the corresponding $X^m = \zeta_I X^I$ to zero, obtaining an {\it effective} prepotential
\be \label{blockmabjm}
F\,=\,-2 i\, \sqrt{X^0 X^1 X^2 X^3} \quad \Rightarrow \quad
F^\text{eff}=\, -2 i\, \sqrt{ (X^1+X^2+X^3) X^1 X^2 X^3}\,,
\ee
which can be used in the gravitational blocks.

We first evaluate the three-sphere free energy in this model, \eqref{eq:S3freenergy}, which gives
\be
	\cF^\text{mABJM}_{S^3} (\varphi) = \frac{\pi}{G_N^{(4)}}\, \sqrt{(\varphi^1+\varphi^2+\varphi^3) \varphi^1 \varphi^2 \varphi^3}\ ,
\ee 
under the constraint $\sum_{i=1}^3 \varphi^i = 1$. It is easy to extremize the above formula, finding simply $\bar \varphi^i = 1/3$ for all $i$, which corresponds to the superconformal point of massive ABJM theory. We therefore find 
\begin{equation} \label{b3abjm}
\cF^\text{mABJM}_{S^3} := \cF^\text{mABJM}_{S^3} (\bar \varphi) = \frac{\pi}{3 \sqrt{3}\, G^{(4)}_N} = \frac{\pi\, L^2_{AdS_4}}{2\, G^{(4)}_N}\,,
\end{equation}
as anticipated in \eqref{flads4}. Since the ABJM and mABJM models come from compactification on the same manifold, $S^7$, it follows that their respective Newton constants are equal, such that
\be
	\cF^\text{mABJM}_{S^3} = \frac{4}{3 \sqrt{3}}\, \cF^\text{ABJM}_{S^3}\ .
\ee
We have thus established the well-known relation between the free energies of ABJM and massive ABJM from the gravitational block picture.

Now we consider the spindle entropy function, given by \eqref{eq:entropyextrem}. As we have $\frak{n}^m = \zeta_I \frak{n}^I = 0$ such that $\mathfrak{n}^0+\mathfrak{n}^1+\mathfrak{n}^2+\mathfrak{n}^3=2(\mathfrak{n}^1+\mathfrak{n}^2+\mathfrak{n}^3)$, the twist condition \eqref{blocktwist} reduces to 
\begin{equation} \label{nconst111}
\frak{n}^0 = \sum_{i=1}^3\mathfrak{n}^i\,=\,\frac{n_++\sigma{n}_-}{2n_+n_-}\,,
\end{equation} 
and the constraint in \eqref{varphicst} gives
\begin{equation} \label{nconst222}
\sum_{i=1}^3\varphi^i-\frac{n_+-\sigma{n}_-}{2n_+n_-}\epsilon\,=\,1\,.
\end{equation}
We choose $\sigma = -1$ and find for the entropy function
\be
S^- \left(\varphi,\epsilon;\frak{n}, n_+, n_-\right)\,=\,  -\frac{i\, \pi}{8\, G^{(4)}_N\, \epsilon}\, \Big( F^\text{eff}\left(\varphi^i + \epsilon \frak{n}^I \right) + F^\text{eff}\left(\varphi^i - \epsilon \frak{n}^I \right)  \Big)\ ,
\ee
using \eqref{blockmabjm}.
 Extremizing this entropy function with the constraints in \eqref{nconst111} and \eqref{nconst222}, we can obtain the Bekenstein-Hawking entropy. Due to the complexity of equations we could not obtain analytic expression. However, any set of specific choices of spindle numbers and fluxes allows for a direct numeric solutions that can be readily compared with the explicit solutions and successfully matched, see the attached {\it Mathematica} file. The precise match requires the following identification of the parameters in \cite{Suh:2022pkg},
\begin{equation} \label{paraetf}
n_N\,=\,n_+\,, \quad n_S\,=\,n_-\,, \quad \frac{p_{F_1}}{n_+n_-}\,=\,-\frac{1}{2}\left(\mathfrak{n}^1-\mathfrak{n}^2\right)\,, \quad \frac{p_{F_2}}{n_+n_-}\,=\,-\frac{1}{\sqrt{3}}\left(\mathfrak{n}^1+\mathfrak{n}^2-2\mathfrak{n}^3\right)\,.
\end{equation}

Furthermore, the value of the scalar fields at the poles can be matched with the extremal values of the fugacities,
\begin{align}
\frac{\bar \varphi^1+\mathfrak{n}^1\, \bar \epsilon}{1+\bar \epsilon/n_+}\,=\,e^{-\lambda_{2}-\lambda_{3}} |_N\,, \notag \\
\frac{\bar \varphi^1-\mathfrak{n}^1\, \bar\epsilon}{1+\bar \epsilon/n_-}\,=\,e^{-\lambda_{2}-\lambda_{3}} |_S\,,
\end{align}
and permutations of $\{1, 2, 3 \}$.

\subsection{$Q^{1,1,1}$ and $M^{1,1,1}$}
In the case of the $Q^{1,1,1}$ we discussed at length here, we start with the same prepotential as in ABJM theory, 
\begin{equation}
F = - 2 i\, \sqrt{X^0 X^1 X^2 X^3}\ ,
\end{equation} 
but a different R-symmetry and massive vector fields
\begin{equation}
A^R\,=\,4 \sqrt{2}A^0\,, \qquad
A^m =\sqrt{2}\, \left( e_0 A^0 - 2 A^1 - 2 A^2 - 2 A^3 \right)\ ,
\end{equation}
that corresponds to $\xi_0 = 4 \sqrt{2}, \xi_1 = \xi_2 = \xi_3 = 0$, and $\zeta_0 =\sqrt{2}\, e_0, \zeta_1 = \zeta_2= \zeta_3 = - 2 \sqrt{2}$ in the notation of the introductory section. As discussed in section \ref{sec:2}, the AdS length scale in this case is given by $L_{AdS_4}^2 = \sqrt{e_0^3}/4 \sqrt{6^3}$.

Again, we can directly apply the massive vector constraint on the level of the prepotential, setting the corresponding $X^m = \zeta_I X^I$ to zero, obtaining an {\it effective} prepotential
\be \label{blockmq111}
F\,=\,-2 i\, \sqrt{X^0 X^1 X^2 X^3} \quad \Rightarrow \quad
F^\text{eff}=\, -2 i\, \sqrt{\tfrac2{e_0}(X^1+X^2+X^3) X^1 X^2 X^3 }\,,
\ee
which can be used in the gravitational blocks.

We first evaluate the three-sphere free energy in this model, \eqref{eq:S3freenergy}, which gives
\be
	\cF^{Q^{1,1,1}}_{S^3} (\varphi) = \frac{\pi}{G_N^{(4)}}\, \sqrt{\tfrac2{e_0}(\varphi^1+\varphi^2+\varphi^3) \varphi^1 \varphi^2 \varphi^3}\ ,
\ee 
under the constraint $\varphi^0 = 2/e_0\, \sum_{i=1}^3 \varphi^i = 1/(2\sqrt{2})$. It is easy to extremize the above formula, finding simply $\bar \varphi^i = e_0/(12\, \sqrt{2})$ for all $i$, which corresponds to the superconformal point of the flavoured ABJM theory. We therefore find
\begin{equation}
\cF^{Q^{1,1,1}}_{S^3} : = \cF^{Q^{1,1,1}}_{S^3} (\bar \varphi) = \frac{\pi\, \sqrt{e_0^3}}{8\, \sqrt{6^3}\, G_N^{(4)}}  = \frac{\pi\, L^2_{AdS_4}}{2\, G^{(4)}_N}\, ,
\end{equation}
as anticipated in \eqref{flads4}. This in turn relates to the gauge group rank of the flavoured ABJM via \eqref{fabjm}.

Now we consider the spindle entropy function, given by \eqref{eq:entropyextrem}. Since $\frak{n}^m = \zeta_I \frak{n}^I = 0$, the twist condition \eqref{blocktwist} reduces to
\begin{equation} \label{nconst1111}
\mathfrak{n}^0\,=\,\frac{1}{4\sqrt{2}}\frac{n_++\sigma{n}_-}{n_+n_-}\,, \qquad \sum_{i=1}^3\mathfrak{n}^i\,=\,\frac{e_0}{2}\frac{1}{4\sqrt{2}}\frac{n_++\sigma{n}_-}{n_+n_-}\,.
\end{equation}
and the constraint in \eqref{varphicst} gives
\begin{equation} \label{nconst2222}
\varphi^0\,=\,\frac{1}{2\sqrt{2}}+\frac{1}{4\sqrt{2}}\frac{n_+-\sigma{n}_-}{n_+n_-}\epsilon\,, \qquad \sum_{i=1}^3\varphi^i\,=\, \frac{e_0}{2}\left(\frac{1}{2\sqrt{2}}+\frac{1}{4\sqrt{2}}\frac{n_+-\sigma{n}_-}{n_+n_-}\epsilon\right)\,,
\end{equation}
We choose $\sigma = -1$ and find for the entropy function
\be
S^- \left(\varphi,\epsilon;\frak{n}, n_+, n_-\right)\,=\,  -\frac{i\, \pi}{8\, G^{(4)}_N\, \epsilon}\,  \Big( F^\text{eff}\left(\varphi^i + \epsilon \frak{n}^I \right) + F^\text{eff}\left(\varphi^i - \epsilon \frak{n}^I \right)  \Big)\ ,
\ee
using \eqref{blockmq111}.
Extremizing this entropy function with the constraints in \eqref{nconst1111} and \eqref{nconst2222}, we can obtain the Bekenstein-Hawking entropy. Again we could not obtain analytic expression of the on-shell entropy, but made sure to numerically match the results. For specific choices of spindle numbers and fluxes with the identification of parameters reflecting \eqref{vectorsrmbb},
\begin{equation}
n_N\,=\,n_-\,, \quad n_S\,=\,n_+\,, \quad \frac{p_{\mathcal{B}_1}}{n_+n_-}\,=\, 4\sqrt{2}\left(\mathfrak{n}^1-2\mathfrak{n}^2+\mathfrak{n}^3\right)\,, \quad \frac{p_{\mathcal{B}_2}}{n_+n_-}\,=\, 4\sqrt{2}\left(\mathfrak{n}^1-\mathfrak{n}^3\right)\,,
\end{equation}
the Bekenstein-Hawking entropy numerically matches the result obtained from the solution in \eqref{bhent}. This can be again seen explicitly in the attached {\it Mathematica} file.

As before, the value of the scalar fields at the poles can be matched with the extremal values of the fugacities,
\begin{align}
\frac{4\sqrt{2}}{e_0}\frac{\bar \varphi^1+\mathfrak{n}^1\, \bar\epsilon}{1+\bar \epsilon/n_-}\,=\,e^{u_{2}+u_{3}} |_N\,, \notag \\
\frac{4\sqrt{2}}{e_0}\frac{\bar \varphi^1-\mathfrak{n}^1\, \bar \epsilon}{1+\bar \epsilon/n_+}\,=\,e^{u_{2}+u_{3}} |_S\,,
\end{align}
and permutation of indices $\{1, 2, 3\}$.

Calculating the Bekenstein-Hawking entropy of spindle black holes in $AdS_4\times{M}^{1,1,1}$ from the gravitational blocks readily follows from the above calculation in parallel by setting
\begin{equation}
X^3\,=\, X^2\,, \qquad \mathfrak{n}^3\,=\,\mathfrak{n}^2\,, \qquad \varphi^3\,=\,\varphi^2\,.
\end{equation}
For specific choices of spindle numbers and fluxes, the result numerically matches the Bekenstein-Hawking entropy obtained from the solution in \eqref{bhent}.

\section{Discussion}
\label{sec:7}
There are a number of open questions and ways to extend the present work, which we hope to explore in future. We list some of them below.
\begin{itemize}

	\item As evident from the gravitational block form and the explicit solution ansatz, we have allowed for both twist and anti-twist solutions to the models of interest. Unlike the anti-twist case, on which we focused most of the discussion, we were not able to find consistent solutions with positive entropy and scalars in the twist class. However, we were still able to test (in the numeric approach in the attached {\it Mathematica} file) that the on-shell answers from gravitational blocks agree with the explicit numeric solutions. Importantly, the gravitational blocks should also hold {\it off-shell} without any extremization, where they correspond to more general Euclidean saddles with no Lorentzian analog, see e.g.\ \cite{Bobev:2020pjk}. Our checks therefore hint further at the strong expectation that our main results, \eqref{eq:1}-\eqref{eq:3}, hold equally well with both twist and anti-twist.

	\item A natural generalization of the present results is to include electric charges and non-vanishing angular momentum to the near-horizon solutions we have discovered. The gravitational block picture, \eqref{eq:1}-\eqref{eq:3}, gives a clear prediction of how the entropy and scalars should change with the addition of extra charges, but it would be desirable to prove this directly from the BPS equations after a more general spacetime ansatz is made. A related problem, also anticipated by the gravitational block construction by setting $n_+ = n_- = 1$ in \eqref{eq:1}-\eqref{eq:3}, is the search for rotating twisted and non-twisted (or {\it anti-twisted} in the present language) spherical black holes that would generalize the results of \cite{Hristov:2018spe} and \cite{Hristov:2019mqp}, respectively, to the case of charged hypermultiplet scalars and massive vectors extending the static twisted solutions of \cite{Halmagyi:2013sla}.

	\item Another generalization is to enlarge the class of supergravity models in order to allow for general dyonic gauging, such as the one coming from compactifications of massive type IIA supergravity, \cite{Guarino:2015jca}. We expect the dyonic gauging to give rise to constraints in the gravitational blocks similar to the ones in this work based on an effective form of the prepotential, as already implied by the twisted black holes solutions in \cite{Hosseini:2017fjo, Benini:2017oxt}.

	\item  It would be interesting to write down all present results in the language of symplectic invariant and covariant quantities as in \cite{Hristov:2018spe,Hristov:2019mqp}, which was partially done in \cite{Couzens:2021rlk} for spindles in STU supergravity. The manifest symplectic covariant form of the gravitational blocks was instead developed in \cite{Hosseini:2023ewi}, and we expect our results to also fit inside this framework. The possible upshot in this approach is the higher likelihood of finding precise analytic structures for the present solutions.

	\item When the solutions we construct are uplifted to eleven-dimensional supergravity, it will fall in the class of GK geometry, \cite{Kim:2006qu, Gauntlett:2007ts}. Thus we should note that a complementary point of view towards the entropy function in 4d is the idea of volume extremization of the internal manifold, put forward in e.g. \cite{Couzens:2018wnk, Hosseini:2019ddy, Gauntlett:2019roi,Kim:2019umc} for black holes and black strings. In particular, one can translate also the gravitational block idea in terms of the internal volume, see \cite{Boido:2022mbe,Martelli:2023oqk}. This has the potential upshot of removing the need for lower-dimensional truncations. It would be interesting to reproduce the present results directly working with the SE$_7$ data.

         \item A related comment concerns the implicit shortcoming of our 4d approach as relying on the existing truncations of \cite{Cassani:2012pj}. These truncations are aimed at keeping the modes from non-trivial two/five-cycles such that they preserve the so-called baryonic symmetries, but unfortunately do not keep any of the existing mesonic symmetries associated with the isometry group. This is particularly unfortunate for the examples of $V^{5,2}$ and $N^{0,1,0}$ manifolds that have no baryonic symmetries and consequently one can only embed minimal supergravity solutions in these models,~\footnote{See, for example, black hole solutions with the horizons of $AdS_2\times\Sigma_\mathfrak{g}$ in this model where $\Sigma_\mathfrak{g}$ is a Riemann surface with genus, $\mathfrak{g}$. in \cite{Erbin:2014hsa,Azzurli:2017kxo} and their field theory description in \cite{Hosseini:2016tor, Hosseini:2016ume}. It is automatic that the spindle solutions in minimal supergravity presented here, see sections \ref{sec41} and \ref{sec:61}, hold in these cases.} missing out on the detailed internal structure of these cases. The same problem is evaded for the example of $S^7$ (mABJM) where we know the mesonic symmetries due to the existing truncation to maximal $\cN=8$ gauged supergravity in 4d. Similarly, it would be interesting to generalize the truncations of \cite{Cassani:2012pj} in order to account for the additional symmetries, see e.g. \cite{Hosseini:2019ddy}.

\end{itemize}

\bigskip

\leftline{\bf Acknowledgements}
\noindent We would like to thank Seyed Morteza Hosseini, Hyojoong Kim and Chiara Toldo for useful discussions. The study of KH is financed by the European Union- NextGenerationEU, through the National Recovery and Resilience Plan of the Republic of Bulgaria, project No BG-RRP-2.004-0008-C01. MS was supported by the Kavli Institute for Theoretical Sciences (KITS) and the University of Chinese Academy of Sciences (UCAS).


\appendix
\section{Consistent truncation of M-theory on $Q^{1,1,1}$} \label{appA}
\renewcommand{\theequation}{A.\arabic{equation}}
\setcounter{equation}{0} 

\subsection{The $\mathcal{N}=2$ formalism} \label{appA11}

The consistent truncation of eleven-dimensional supergravity, \cite{Cremmer:1978km}, on seven-dimensional Sasaki-Einstein manifolds was performed in \cite{Cassani:2012pj}. In particular, we consider the seven-dimensional Sasaki-Einstein manifold, $Q^{1,1,1,}$, which is a coset space, $\frac{SU(2)\times{SU}(2)\times{SU}(2)}{U(1)\times{U}(1)}$. It has two non-trivial two-cycles and the dimensionally reduced theory contains two Betti vector multiplets. The field theory dual to the $AdS_4\times{Q}^{1,1,1}$ solutions is 3d flavored ABJM theories, \cite{Benini:2009qs, Cremonesi:2010ae}. The field content at the $\cN=2$ vacuum is as follows, \cite{Cassani:2012pj, Monten:2021som},
\begin{itemize}
\item The gravity multiplet contains the metric and the graviphoton,

\item A massive vector multiplet contains a massive vector field with $m^2l_{AdS}^2=12$, dual to an operator with $\Delta=5$, which has eaten the axion, $\sigma$. There are also five scalar fields with $m^2l_{AdS_4}^2=(18,10,10,10,4)$ dual to operators with $\Delta=(6,5,5,5,4)$.

\item Two Betti vector multiplets: each contains a massless vector field and a complex scalar field with $m^2l_{AdS_4}^2=-2>m_\text{BF}$ dual to operators with either $\Delta=2,1$ by the choice of boundary conditions.
\end{itemize}

The complete 4d $\mathcal{N}\,=\,2$ truncation on $Q^{1,1,1}$ in supergravity language consists of the gravity multiplet with the aforementioned graviton and graviphoton, $\{g_{\mu\nu}, A^0_\mu\}$, 3 vector multiplets with a vector and complex scalar, $\{A^i_\mu, z^i\}$, and 1 hypermultiplet (known as the {\it universal} hypermultiplet) with four real scalars, $\{\phi, \sigma, \xi^0, \tilde{\xi}_0\}$. The scalar fields from the vector multiplets and the hypermultiplet parametrize the coset manifolds,
\begin{equation}
\mathcal{M}_v\times\mathcal{M}_h\,=\,\left(\frac{SU(1,1)}{U(1)}\right)^3\,\times\,\frac{SU(2,1)}{S(U(2)\,\times\,U(1))}\,,
\end{equation}
which is a product of special K\"ahler and quaternionic manifolds, respectively.

We move to discuss in more detail these two scalar manifolds that play an important role in writing down the BPS variations.

{\bf Universal hypermultiplet:} Here we gather the relevant quantities and specific gaugings of the universal hypermultiplet, given by the metric $\mathcal{M}_h$. Written in terms of real coordinates, $\{\phi,\,\sigma,\,\xi^0,\,\tilde{\xi}_0\}$, the metric is  
\begingroup
\renewcommand*{\arraystretch}{1.2}
\begin{equation}
h = 
\begin{pmatrix}
1 & 0 & 0 & 0 \\
0 & \frac{1}{4} e^{4\phi} & - \frac{1}{8} e^{4\phi} \tilde{\xi}_0 & \frac{1}{8} e^{4\phi} \xi^0 \\
0 & - \frac{1}{8} e^{4\phi} \tilde{\xi}_0\, & \, \frac{1}{4} e^{2\phi}\left(1 + \frac{1}{4} e^{2\phi}(\tilde{\xi}_0)^2\right) & -\frac{1}{16} e^{4\phi} \xi^0 \tilde{\xi}_0 \\
0 &  \frac{1}{8} e^{4\phi} \xi^0 & -\frac{1}{16} e^{4\phi} \xi^0 \tilde{\xi}_0 &  \frac{1}{4} e^{2\phi}\left(1 + \frac{1}{4} e^{2\phi} \left(\xi^0\right)^2\right)
\end{pmatrix}.
\end{equation}
\endgroup
The isometry group, $SU(2,1)$, has eight generators; two of these are used for gauging in the model we consider explicitly below, generating the group, $\mathbb{R}\times{U}(1)$.~\footnote{See e.g.\ \cite{Halmagyi:2011xh} for a careful discussion of the isometries and the physical outcome of their gauging.} The corresponding Killing vectors are
\begin{equation}
k^\mathbb{R}\,= \,\partial_\sigma\,, \qquad k^{U(1)}\,=\,-\tilde{\xi}_0\partial_{\xi^0}+\xi^0\partial_{\tilde{\xi}_0}\ .
\end{equation}
These two isometries are gauged by a particular linear combination of the vector fields in the theory. One defines Killing vectors with a symplectic index corresponding to each of the full set of electric and magnetic gauge fields at our disposal. The moment maps associated to these two Killing vectors are
\begin{equation}
P^\mathbb{R}\,=\,\left(0,\,0,\,-\frac{1}{2}e^{2\phi}\right), \qquad P^{U(1)}\,=\,\left(\tilde{\xi}_0e^\phi,\,-\xi^0e^\phi,\,1-\frac{1}{4}\left((\xi^0)^2+(\tilde{\xi}_0)^2\right)e^{2\phi}\right)\ . 
\end{equation}

In order to make sure the moment maps are strictly in the third direction, we can set $\xi^0\,=\,\tilde{\xi}_0\,=\,0$ guaranteeing that $k^{U(1)} = 0$ independent of the details of the scalar manifold for the vector multiplets. On the contrary, $k^\mathbb{R}\neq0$ always and thus we would find a genuine constraint on the vector multiplets from this type of gauging. Now the moment map $P$ remains non-zero only along the third direction, 
\begin{equation}
P^\mathbb{R}\,=\,\left(0,\,0,\,-\frac{1}{2}e^{2\phi}\right)\,, \qquad P^{U(1)}(\xi^0\,=\,\tilde{\xi}_0\,=\,0)\,=\,\left(0,\,0,\,1\right)\,.
\end{equation}
From now on we will only discuss this third, or $z$, component of the moment maps. The choice of setting  $\xi^0\,=\,\tilde{\xi}_0\,=\,0$ is not in itself a subtruncation to a smaller $\cN=2$ supergravity, but it can always be made on a given background without breaking further supersymmetry.

Note also that this way of solving the hyperscalar equations for the universal hypermultiplet, by setting $\xi^0=\tilde{\xi}_0=0$ and keeping only $k^\sigma$ non-vanishing also means that the $SU(2)$ connection takes a simple form, 
\begin{equation}
\omega^x\,=\,e^\phi{\rm d}\xi^0\,=\,0\,, \qquad \omega^y\,=\,e^\phi{\rm d}\tilde{\xi}_0\,=\,0\,, \qquad \omega^z\,=\,-\frac{1}{2}e^{2\phi}\left({\rm d}\sigma+\xi^0{\rm d}\tilde{\xi}_0\right)\,=\,-\frac{1}{2}e^{2\phi}{\rm d}\sigma\,.
\end{equation}
The relevant part of the corresponding $SU(2)$ curvature, defined as $\Omega^x := {\rm d} \omega^x - \tfrac12 \epsilon^{x y z} \omega^y \wedge \omega^z$, is therefore
\be
	\Omega^z_{\sigma \phi} = - \Omega^z_{\phi \sigma} =\frac12 e^{2\phi}\, .
\ee

{\bf Vector multiplets:} The full model is further specified by the vector multiplet geometry, which is given by the so-called STU model, $i.e.$, the coset space $\mathcal{M}_v$ above with a prepotential given in \eqref{eq:prepotential} that defines (see below) the corresponding metric. In order to simplify the model from the start, we assume that there are no axions, such that the three complex scalar fields, $z^{1,2,3}$, are real. The Lagrangian and supersymmetry variations then follow from the choice of holomorphic sections,
\begin{equation}
X^0\,=\,1\,, \qquad X^1\,=\,z^2z^3\,, \qquad X^2\,=\,z^3z^1\,, \qquad X^3\,=\,z^1z^2\,,
\end{equation}
leading to a K\"ahler potential,
\begin{equation}
e^{-K}\,=\,8\text{Re}[z^1]\text{Re}[z^2]\text{Re}[z^3]\,=\,8z^1z^2z^3\,,
\end{equation}
with quantities,
\begin{equation}
K_i\,=\,\partial_{z^i}K\,=\,-\left(2z^i\right)^{-1}\,, \qquad g_{i\bar{j}}\,=\,K_{ij}\,=\,\delta_{ij}\left(2z^i\right)^{-2}\,.
\end{equation}
The so-called period matrix, which defines the gauge field couplings, is given by
\begin{equation}
\mathcal{N}\,=\,-i\, \text{diag}\left(z^1z^2z^3,\,\frac{z^1}{z^2z^3},\,\frac{z^2}{z^3z^1},\,\frac{z^3}{z^1z^2}\right)\,,
\end{equation}
such that $\text{Re}\mathcal{N}=0$. 

{\bf Gauging:} The consistent truncation with universal hypermultiplet gauging coming from the compactification of M-theory on $Q^{1,1,1}$, \cite{Cassani:2012pj}, features a mixed dyonic gaugings. For simplicity we directly consider the consequent symplectic rotation to purely electric gauging (and prepotential which we already anticipated to be \eqref{eq:prepotential}), as presented in \cite{Halmagyi:2013sla}. We have a hypermultiplet gauging,
\begin{equation}
k_I\,=\,\left(\sqrt{2}\left(4\, k^{U(1)}+e_0k^\mathbb{R}\right),\,- 2\sqrt{2}k^\mathbb{R},\, -2\sqrt{2}k^\mathbb{R},\, -2\sqrt{2}k^\mathbb{R}\right)\,,
\end{equation}
with the constant, $e_0$, an unfixed Freund-Rubin parameter, and thus we have
\begin{equation}
P_I^z\,=\,\left(\sqrt{2}\left(4-\frac{1}{2}e^{2\phi}e_0\right),\,e^{2\phi},\,e^{2\phi},\,e^{2\phi}\right)\,.
\end{equation}

{\bf Lagrangian and supersymmetry variations:} Following the conventions of \cite{Andrianopoli:1996cm}, the Lagrangian, after the simplifications of taking real vector multiplet scalars and $\xi^0\,=\,\tilde{\xi}_0\,=\,0$ reads~{\footnote{Only in this appendix \ref{appA11}, we employ the mostly minus signature and stick to the notation and conventions in \cite{Andrianopoli:1996cm}.}}
\begin{equation}
e^{-1}\mathcal{L}\,=\,\frac{1}{2}R+\sum_i\frac{\left(\partial{z}^i\right)^2}{4\left(z^i\right)^2}+\left(\partial\phi\right)^2+\frac{1}{4}e^{4\phi}\left(D \sigma\right)^2+\text{Im}\mathcal{N}_{IJ}F^I_{\mu\nu}F^{I\mu\nu}-g^2\mathcal{V}\,,
\end{equation}
with the gauge covariant derivative,
\begin{equation}
\label{eq:Dsigma}
D_\mu\sigma\,=\,\partial_\mu\sigma+gA^m_\mu\,,
\end{equation}
where we define the massive vector,
\begin{equation}
A^m\,=\,\sqrt{2}\left(e_0A^0-2 A^1-2 A^2 -2 A^3\right)\,,
\end{equation}
and $I,J=0,\ldots,3$. The scalar potential follows straightforwardly from the data given above, and is discussed further below. The R-symmetry gauge field is essentially chosen by the orientation of the Killing vector $k^{U(1)}$,
\begin{equation}
A^R\,=\,4 \sqrt{2}A^0\,.
\end{equation}
The supersymmetry variations of gravitino, gaugino and hyperino are given by, respectively,
\begin{align} \label{gravitino0}
\delta\psi_{\mu{A}}\,=&\,\nabla_\mu\varepsilon_A+\frac{i}{2}\left(4\sqrt{2}gA^0_\mu-\frac{1}{2}e^{2\phi}\nabla_\mu\sigma\right)\sigma^3{}_A{}^B\varepsilon_B-\frac{g}{2}e^{K/2}P_I^3X^I\gamma_\mu\sigma^3{}_{AB}\varepsilon^B \notag \\ 
&+2ie^{K/2}X^I\text{Im}\mathcal{N}_{IJ}F^{-J}_{\mu\nu}\gamma^\nu\epsilon_{AB}\varepsilon^B\,, \\ \label{gaugino0}
\delta\lambda^{iA}\,=&\,i\partial_\mu{z}^i\gamma^\mu\varepsilon^A+ige^{K/2}g^{ij}\left(\partial_j+K_j\right)X^IP^3_I\sigma^{3AB}\varepsilon_B \notag \\ 
&-e^{K/2}g^{ij}\left(\partial_j+K_j\right)X^I\text{Im}\mathcal{N}_{IJ}F^{-J}_{\mu\nu}\gamma^{\mu\nu}\epsilon^{AB}\varepsilon_B\,, \\ \label{hyperino0}
\delta\zeta_\alpha\,=&\,U_{u\alpha{A}}\left(i\nabla_\mu{q}^u\gamma^\mu\varepsilon^A+2ge^{K/2}k_I^uX^I\epsilon^{AB}\varepsilon_B\right)\,,
\end{align}
and we define quantities for later convenience,
\begin{align}
W\,\equiv&\,e^{K/2}P_I^3X^I\,=\,\frac{1}{4\sqrt{z^1z^2z^3}}\left(8-e_0e^{2\phi}+2e^{2\phi}\left(z^1z^2+z^3z^1+z^2z^3\right)\right)\,, \notag \\
H_{\mu\nu}\,\equiv&\,e^{K/2}X^I\text{Im}\mathcal{N}_{IJ}F^{-J}_{\mu\nu}\,=\,-\frac{1}{2\sqrt{2z^1z^2z^3}}\left(z^1z^2z^3F^{-0}_{\mu\nu}+z^1F^{-1}_{\mu\nu}+z^2F^{-2}_{\mu\nu}+z^3F^{-3}_{\mu\nu}\right)\,.
\end{align}

\subsection{Parametrizations and equations of motion}

The three complex scalar fields, $z^i$, are often rewritten using the parametrization,~{\footnote{Note that, due to the overall $i$ factor we have inserted in the prepotential, \eqref{eq:prepotential}, our parametrization is different from the one in \cite{Cassani:2012pj, Halmagyi:2013sla}, $z^i\,=\,b^i+iv^i$. The physical scalars we keep, $v^i$, are however the same.}}
\begin{equation}
z^i\,=\,ib^i+v^i\,,
\end{equation}
and further
\begin{equation}
v^i\,=\,e^{2u_i}\,.
\end{equation}
 We consider the axion free case with $b^i=0$, as explained above.

The bosonic Lagrangian in this parametrization reads~\footnote{Here we revert to the mostly plus signature used in the main body of this work.}
\begin{align} \label{lagapp}
e^{-1}\mathcal{L}\,&=\,\frac{1}{2}R-\sum_{i=1}^3\partial_\mu{}u_i\partial^\mu{}u_i-\partial_\mu\phi\partial^\mu\phi-\frac{1}{4}e^{4\phi}D_\mu\sigma{}D^\mu\sigma-g^2\mathcal{V} \notag \\
&-\Big[e^{2u_1+2u_2+2u_3}F^0_{\mu\nu}F^{0\mu\nu}+e^{2u_1-2u_2-2u_3}F^1_{\mu\nu}F^{1\mu\nu} \notag \\
& \,\,\,\, +e^{-2u_1+2u_2-2u_3}F^2_{\mu\nu}F^{2\mu\nu}+e^{-2u_1-2u_2+2u_3}F^3_{\mu\nu}F^{3\mu\nu}\Big]\,,
\end{align}
with $D \sigma$ as in \eqref{eq:Dsigma}. The scalar potential is
\begin{align}
\mathcal{V}\,=\,-&8e^{2\phi}\left(e^{-2u_1}+e^{-2u_2}+e^{-2u_3}\right)+e^{4\phi}\left(e^{-2u_1+2u_2+2u_3}+e^{2u_1-2u_2+2u_3}+e^{2u_1+2u_2-2u_3}\right) \notag \\
+&\frac{1}{4}e_0^2e^{4\phi}e^{-2u_1-2u_2-2u_3}\,,
\end{align}
and it can be written as
\begin{equation}
\mathcal{V}\,=\,\sum_{i=1}^3\left(\frac{\partial\,W}{\partial\,u_i}\right)^2+\left(\frac{\partial\,W}{\partial\phi}\right)^2-3W^2\,,
\end{equation}
where the superpotential was defined above, and explicitly reads
\begin{equation}
W\,=\,\frac{1}{4}e^{-u_1-u_2-u_3}\left(8-e_0e^{2\phi}+2e^{2\phi}\left(e^{2u_1+2u_2}+e^{2u_2+2u_3}+e^{2u_3+2u_1}\right)\right)\,.
\end{equation}
 We can also parametrize the two Betti vector fields in the same normalization as the R-symmetry and massive vectors, 
\begin{align}
A^{\text{R}}\,=&\,4\sqrt{2}A^0\,, \notag \\
A^{m}\,=&\,\sqrt{2}\left(e_0A^0-2A^1-2A^2-2A^3\right)\,, \notag \\
A^{\mathcal{B}_1}\,=&\,4\sqrt{2}\left(A^1-2A^2+A^3\right)\,, \notag \\
A^{\mathcal{B}_2}\,=&\,4\sqrt{2}\left(A^1-A^3\right)\,.
\end{align}

The supersymmetry variations of fermionic fields, gravitino, gaugino and hyperino, \eqref{gravitino0}, \eqref{gaugino0}, and \eqref{hyperino0}, reduce to, respectively,~\footnote{In order to bring the hyperino variation to exhibit a free $SU(2)$ index, we make use of the identity $U_{u \alpha A} U_v^{\alpha B} = \tfrac12 h_{u v} \delta_A{}^B + \tfrac{i}2 \Omega^x_{u v} \sigma^3{}_A{}^B$, see \cite{Ceresole:2001wi}.}~{\footnote{In terms of the complex scalar fields, $z^i$, the gaugino variation reduces to
\begin{equation}
\delta\lambda^{iA}\sim\frac{1}{4\left(z^i\right)^2}\partial_\mu{z}^i\gamma^\mu\varepsilon^A+g\partial_{z^i}W\sigma^{3AB}\varepsilon_B-i\partial_{z^i}H_{\mu\nu}\gamma^{\mu\nu}\epsilon^{AB}\varepsilon_B\,.
\end{equation}
}}
\begin{align}
\delta\psi_{\mu\,A}\,\sim\,&\,2\nabla_\mu\varepsilon_A-iB_\mu\sigma^3{}_A{}^B\varepsilon_B-gW\gamma_\mu\sigma^3{}_{AB}\varepsilon^B+4iH_{\mu\nu}\gamma^\nu\epsilon_{AB}\varepsilon^B\,, \notag \\
\delta\lambda^{iA}\,\sim\,&\,\partial_\mu{}u_i\gamma^\mu\varepsilon^A+g\partial_{u_i}W\sigma^{3AB}\varepsilon_B-i\partial_{u_i}H_{\mu\nu}\gamma^{\mu\nu}\epsilon^{AB}\varepsilon_B\,, \notag \\
U^{\alpha\, A}_\phi\, \delta\zeta_\alpha\,\sim\, &  U^{\alpha\, A}_\sigma\, \delta\zeta_\alpha\,\sim\,\,\partial_\mu\phi\gamma^\mu\varepsilon^A+\frac{i}{2}\partial_\phi{B}_\mu\gamma^\mu \sigma^3{}_B{}^A\varepsilon^B+g\partial_\phi\,W\sigma^{3AB}\varepsilon_B\,,
\end{align}
where we have
\begin{align}
H_{\mu\nu}=&\,-\frac{1}{2\sqrt{2}}\left(e^{u_1+u_2+u_3}F^{-0}_{\mu\nu}+e^{u_1-u_2-u_3}F^{-1}_{\mu\nu}+e^{-u_1+u_2-u_3}F^{-2}_{\mu\nu}+e^{-u_1-u_2+u_3}F^{-3}_{\mu\nu}\right)\,, \notag \\
B_\mu\,=&\,-4\sqrt{2}gA_\mu^0+\frac{1}{2}e^{2\phi}D_\mu\sigma\,.
\end{align}
The anti-self-dual part of the field strengths are given by
\begin{equation}
F^{-I}\,=\,\frac{1}{2}\left(F^I-i*F^I\right)\,.
\end{equation}
We express $F^{-I}$ in terms of $F^{-IJ}$,
\begin{align}
F^{-0}\,=&\,\frac{1}{\sqrt{2}}e^{-u_1-u_2-u_3}\left(\overline{F}^{-12}+\overline{F}^{-34}+\overline{F}^{-56}+\overline{F}^{-78}\right)\,, \notag \\
F^{-1}\,=&\,\frac{1}{\sqrt{2}}e^{-u_1+u_2+u_3}\left(\overline{F}^{-12}-\overline{F}^{-34}-\overline{F}^{-56}+\overline{F}^{-78}\right)\,, \notag \\
F^{-2}\,=&\,\frac{1}{\sqrt{2}}e^{u_1-u_2+u_3}\left(-\overline{F}^{-12}+\overline{F}^{-34}-\overline{F}^{-56}+\overline{F}^{-78}\right)\,, \notag \\
F^{-3}\,=&\,\frac{1}{\sqrt{2}}e^{u_1+u_2-u_3}\left(-\overline{F}^{-12}-\overline{F}^{-34}+\overline{F}^{-56}+\overline{F}^{-78}\right)\,,
\end{align}
where we introduce
\begin{equation}
\overline{F}_{\mu\nu}^{-12}\,=\,-4\partial_{u_1}H_{\mu\nu}\,, \qquad
\overline{F}_{\mu\nu}^{-34}\,=\,-4\partial_{u_2}H_{\mu\nu}\,, \qquad
\overline{F}_{\mu\nu}^{-56}\,=\,-4\partial_{u_3}H_{\mu\nu}\,, \qquad
\overline{F}_{\mu\nu}^{-78}\,=\,4H_{\mu\nu}\,.
\end{equation}

To make a direct connection to the solutions and parametrization in \cite{Suh:2022pkg}, we can also introduce a complex Dirac spinor, $\epsilon$, instead of the Weyl spinors, $\varepsilon^A$, 
\begin{equation}
\epsilon\,=\,\varepsilon_1+\varepsilon^2\,.
\end{equation}
The gravitino, gaugino and hyperino variations reduce to, respectively,
\begin{align}
\Big[2\nabla_\mu-iB_\mu-gW\gamma_\mu+4iH_{\mu\nu}\gamma^\nu\Big]\epsilon\,=&\,0\,, \notag \\
\Big[\partial_\mu{u}_i\gamma^\mu+g\partial_{u_i}W+i\partial_{u_i}H_{\mu\nu}\gamma^{\mu\nu}\Big]\epsilon\,=&\,0\,, \notag \\
\left[\partial_\mu\phi\gamma^\mu+g\partial_\phi{W}+\frac{i}{2}\partial_\phi{B}_\mu\gamma^\mu\right]\epsilon\,=&\,0\,.
\end{align}

We present the equations of motion from the Lagrangian in \eqref{lagapp}. The Einstein equations are
\begin{align}
R_{\mu\nu}-&\frac{1}{2}Rg_{\mu\nu}+g^2\mathcal{V}g_{\mu\nu}-2\left(T_{\mu\nu}^\phi+T_{\mu\nu}^{u_1}+T_{\mu\nu}^{u_2}+T_{\mu\nu}^{u_3}\right)-\frac{1}{2}e^{4\phi}T_{\mu\nu}^\sigma \notag \\
-&e^{2\left(u_1+u_2+u_3\right)}T_{\mu\nu}^{A^0}-e^{2\left(u_1-u_2-u_3\right)}T_{\mu\nu}^{A^1}-e^{-2\left(u_1-u_2+u_3\right)}T_{\mu\nu}^{A^2}-e^{-2\left(u_1+u_2-u_3\right)}T_{\mu\nu}^{A^3}\,=\,0\,,
\end{align}
where the energy-momentum tensors are
\begin{align}
T_{\mu\nu}^X\,=&\,\partial_\mu{X}\partial_\nu{X}-\frac{1}{2}g_{\mu\nu}\partial_\rho{X}\partial^\rho{X}\,, \notag \\
T_{\mu\nu}^{A^I}\,=&\,g^{\rho\sigma}F_{\mu\rho}^I{F}_{\nu\sigma}^I-\frac{1}{4}g_{\mu\nu}F_{\rho\sigma}^I{F}^{I\rho\sigma}\,,
\end{align}
and $X$ denotes a scalar field. The Maxwell equations are
\begin{align}
\partial_\nu\left(\sqrt{-g}e^{2\left(u_1+u_2+u_3\right)}F^{0\mu\nu}\right)+\frac{e_0}{2}\sqrt{2}\sqrt{-g}e^{4\phi}g^{\mu\nu}D_\nu\sigma\,=&\,0\,, \notag \\
\partial_\nu\left(\sqrt{-g}e^{2\left(u_1-u_2-u_3\right)}F^{1\mu\nu}\right)-\sqrt{2}\sqrt{-g}e^{4\phi}g^{\mu\nu}D_\nu\sigma\,=&\,0\,, \notag \\
\partial_\nu\left(\sqrt{-g}e^{-2\left(u_1-u_2+u_3\right)}F^{2\mu\nu}\right)-\sqrt{2}\sqrt{-g}e^{4\phi}g^{\mu\nu}D_\nu\sigma\,=&\,0\,, \notag \\
\partial_\nu\left(\sqrt{-g}e^{-2\left(u_1+u_2-u_3\right)}F^{3\mu\nu}\right)-\sqrt{2}\sqrt{-g}e^{4\phi}g^{\mu\nu}D_\nu\sigma\,=&\,0\,.
\end{align}
The scalar field equations are
\begin{align}
&\frac{1}{\sqrt{-g}}\partial_\mu\left(\sqrt{-g}g^{\mu\nu}\partial_\nu{}u_1\right)-\frac{g^2}{2}\frac{\partial\mathcal{V}}{\partial{}u_1} \notag \\
&+e^{2\left(u_1+u_2+u_3\right)}F_{\mu\nu}^0F^{0\mu\nu}+e^{2\left(u_1-u_2-u_3\right)}F_{\mu\nu}^1F^{1\mu\nu} \notag \\
&-e^{-2\left(u_1-u_2+u_3\right)}F_{\mu\nu}^2F^{2\mu\nu}-e^{-2\left(u_1+u_2-u_3\right)}F_{\mu\nu}^3F^{3\mu\nu}\,=\,0\,, \notag \\  \notag \\
&\frac{1}{\sqrt{-g}}\partial_\mu\left(\sqrt{-g}g^{\mu\nu}\partial_\nu{}u_2\right)-\frac{g^2}{2}\frac{\partial\mathcal{V}}{\partial{}u_2} \notag \\
&+e^{2\left(u_1+u_2+u_3\right)}F_{\mu\nu}^0F^{0\mu\nu}-e^{2\left(u_1-u_2-u_3\right)}F_{\mu\nu}^1F^{1\mu\nu} \notag \\
&+e^{-2\left(u_1-u_2+u_3\right)}F_{\mu\nu}^2F^{2\mu\nu}-e^{-2\left(u_1+u_2-u_3\right)}F_{\mu\nu}^3F^{3\mu\nu}\,=\,0\,, \notag \\  \notag \\
&\frac{1}{\sqrt{-g}}\partial_\mu\left(\sqrt{-g}g^{\mu\nu}\partial_\nu{}u_3\right)-\frac{g^2}{2}\frac{\partial\mathcal{V}}{\partial{}u_3} \notag \\
&+e^{2\left(u_1+u_2+u_3\right)}F_{\mu\nu}^0F^{0\mu\nu}-e^{2\left(u_1-u_2-u_3\right)}F_{\mu\nu}^1F^{1\mu\nu} \notag \\
&-e^{-2\left(u_1-u_2+u_3\right)}F_{\mu\nu}^2F^{2\mu\nu}+e^{-2\left(u_1+u_2-u_3\right)}F_{\mu\nu}^3F^{3\mu\nu}\,=\,0\,,
\end{align}
and
\begin{equation}
\frac{1}{\sqrt{-g}}\partial_\mu\left(\sqrt{-g}g^{\mu\nu}\partial_\nu\phi\right)-\frac{g^2}{2}\frac{\partial\mathcal{V}}{\partial\phi}-\frac{1}{2}e^{4\phi}D_\mu\sigma{D}^\mu\sigma\,=\,0\,.
\end{equation}

\subsection{Truncation to minimal gauged supergravity} \label{appA1}

There is a truncation to minimal gauged supergravity. We have the scalar fields to be at their values of the $AdS_4\times{Q}^{1,1,1}$ vacuum and impose the gauge fields to be
\begin{align}
&e^{2u_i}\,=\,\sqrt{\frac{e_0}{6}}\,, \qquad e^{-2\phi}\,=\,\frac{e_0}{6}\,, \qquad L_{AdS_4}\,=\,\frac{1}{2}\left(\frac{e_0}{6}\right)^{3/4}\,, \notag \\
&A\,\equiv\, A^0\,=\,A^1\,=\,A^2\,=\,A^3\,,
\end{align}
where in the latter equation we set $e_0=6$. We find
\begin{equation}
e^{-1}\mathcal{L}\,=\,\frac{1}{2}R+12g^2-4F_{\mu\nu}F^{\mu\nu}\,,
\end{equation}
where $F=dA$. When we have $g=1/2$ and $F\rightarrow1/\left(2\sqrt{2}\right)F$, it reduces to the action of minimal gauged supergravity with the normalization employed in \cite{Ferrero:2020twa}.

\section{Derivation of the BPS equations} \label{appB}
\renewcommand{\theequation}{B.\arabic{equation}}

We consider the metric and the gauge fields,
\begin{align}
ds^2\,=&\,e^{2V}ds_{AdS_2}^2+f^2dy^2+h^2dz^2\,, \notag \\
A^I\,=&\,a^Idz\,,
\end{align}
where $V$, $f$, $h$, and $a^I$, $I=0,\ldots,3$, are functions of coordinate, $y$. The gamma matrices are chosen to be
\begin{equation}
\gamma^m\,=\,\Gamma^m\otimes\sigma^3\,, \qquad \gamma^2\,=\,1_2\otimes\sigma^1\,, \qquad \gamma^3\,=\,1_2\otimes\sigma^2\,,
\end{equation}
and the spinors,
\begin{equation}
\epsilon\,=\,\psi\otimes\chi\,,
\end{equation}
where $\Gamma^m$ are two-dimensional gamma matrices. The two-dimensional spinor satisfies
\begin{equation}
D_m\psi\,=\,\frac{1}{2}\kappa\Gamma_m\psi\,,
\end{equation}
where $\kappa=\pm1$.

For the $AdS_2$ directions, the gravitino variations reduce to
\begin{equation}
\left[-i\left(\kappa{e}^{-V}+4H_{23}\right)\gamma^{23}+V'f^{-1}\gamma^2\right]\epsilon\,=\,gW\epsilon\,.
\end{equation}
It reduces to a projection condition,
\begin{equation}
\label{eq:projection}
\left[i\cos\xi\gamma^{23}+\sin\xi\gamma^2\right]\epsilon\,=\,\epsilon\,,
\end{equation}
where $\xi$ is introduced, 
\begin{equation} \label{fromxi}
-\kappa{e}^{-V}-4H_{23}\,=\,gW\cos\xi\,, \qquad V'f^{-1}\,=\,gW\cos\xi\,.
\end{equation}
A solution of the projection condition is
\begin{equation}
\epsilon\,=\,e^{i\frac{\xi}{2}\gamma^3}\eta\,, \qquad \gamma^2\eta\,=\,i\gamma^3\eta\,.
\end{equation}
From \eqref{fromxi} we have $\partial_z\xi=0$.  The spinors have definite chirality with respect to $\gamma^{23}$ at $\xi=0,\pi$,
\begin{equation}
\xi\,=\,0\,\pi\,, \qquad \gamma^{23}\epsilon\,=\,\pm{i}\epsilon\,.
\end{equation}

For the $y$ direction, the gravitino variation reduces to
\begin{equation} \label{b10}
\left[\partial_y\eta-\frac{1}{2}V'\eta+\frac{i}{2}\left(\partial_y\xi+8fH_{23}+\kappa{f}e^{-V}\right)\gamma^3\right]\eta\,=\,0\,,
\end{equation}
where \eqref{fromxi} was employed. For the $z$ direction, we have
\begin{align} \label{b11}
\Big[2\partial_z-iB_z+&if^{-1}h'\cos\xi-4iH_{23}h\sin\xi \notag \\
+&\Big(f^{-1}h'\sin\xi-gWh+4H_{23}h\cos\xi\Big)\gamma^3\Big]\eta\,=\,0\,.
\end{align}

We require $a_1^2+a_2^2=0$ if we have $\left(a_1+ia_2\gamma^3\right)\eta=0$. Thus from \eqref{b10} and \eqref{b11} we find
\begin{equation}
\eta\,=\,e^{V/2}e^{isz}\eta_0\,,
\end{equation}
where $\eta_0$ is independent of $y$, $z$, and $s$ is constant. We find
\begin{align}
\partial_y\xi+8fH_{23}+\kappa{f}e^{-V}\,&=\,0\,, \notag \\
\left(s-B_z\right)+f^{-1}h'\cos\xi-4H_{23}h\sin\xi\,&=\,0\,, \notag \\
f^{-1}h'\sin\xi-gWh+4H_{23}h\cos\xi\,&=\,0\,.
\end{align}
Hence, we obtain
\begin{align}
f^{-1}h'\,=&\,gWh\sin\xi-\left(s-B_z\right)\cos\xi\,, \notag \\
hH_{23}\,=&\,\frac{1}{4}gWh\cos\xi+\frac{1}{4}\left(s-B_z\right)\sin\xi\,,
\end{align}
and, from the first equation in \eqref{fromxi}, we find
\begin{equation}
\left(s-B_z\right)\sin\xi\,=\,-2gWh\cos\xi-\kappa{h}e^{-V}\,,
\end{equation}
and hence
\begin{align}
H_{23}\,=&\,-\frac{1}{4}gW\cos\xi-\frac{1}{4}\kappa{e}^{-V}\,, \notag \\
f^{-1}\partial_y\xi\,=&\,2gW\cos\xi+\kappa{e}^{-V}\,.
\end{align}
For $\xi\ne0$ we solve for $\left(s-B_z\right)$ and find
\begin{equation}
f^{-1}\frac{h'}{h}\sin\xi\,=\,\kappa{e}^{-V}\cos\xi+gW\left(1+\cos^2\xi\right)\,.
\end{equation}

The gaugino variation reduces, in a similar manner, to
\begin{equation}
f^{-1}u_i'+g\partial_{u_i}W\sin\xi\,=\,0\,,
\end{equation}
and
\begin{equation}
g\partial_{u_i}W\cos\xi+4\partial_{u_i}H_{23}\,=\,0\,.
\end{equation}

The hyperino variation reduces to
\begin{align}
f^{-1}\phi'\sin\xi+g\partial_\phi{W}\,=&\,0\,, \notag \\
g\partial_\phi{W}\cos\xi-\frac{1}{2}\partial_\phi{B}_z\sin\xi{h}^{-1}\,=&\,0\,.
\end{align}

{\bf Summary:} For $\xi\ne0$, the complete BPS equations are given by
\begin{align}
f^{-1}\xi'\,=&\,2gW\cos\xi+\kappa{e}^{-V}\,, \notag \\
f^{-1}V'\,=&\,gW\sin\xi\,, \notag \\
f^{-1}u_i'\,=&\,-g\partial_{u_i}W\sin\xi\,, \notag \\
f^{-1}\phi'\,=&\,-\frac{g\partial_\phi{W}}{\sin\xi}\,, \notag \\
f^{-1}\frac{h'}{h}\sin\xi\,=&\,\kappa{e}^{-V}\cos\xi+gW\left(1+\cos^2\xi\right)\,,
\end{align}
with two constraints,
\begin{align}
\left(s-B_z\right)\sin\xi\,=&\,-2gWh\cos\xi-\kappa{h}e^{-V}\,, \notag \\
g\partial_\phi{W}\cos\xi\,=&\,\frac{1}{2}\partial_\phi{B}_z\sin\xi{h}^{-1}\,.
\end{align}
The field strengths of gauge fields are given by
\begin{align}
\partial_{u_i}H_{23}\,=&\,-\frac{1}{4}g\partial_{u_i}W\cos\xi\,, \notag \\
H_{23}\,=&\,-\frac{1}{4}gW\cos\xi-\frac{1}{4}\kappa{e}^{-V}\,.
\end{align}
The BPS equations are consistent with the equations of motion from the Lagrangian.

There is also a relation,
\begin{equation}
\partial_yW\,=\,-gf\sin\xi\left[\sum_{i=1}^3\left(\partial_{u_i}W\right)^2+\frac{1}{\sin^2\xi}\left(\partial_\phi{W}\right)^2\right]\,,
\end{equation}
and the superpotential is monotonic along the BPS flow if the sign of $f\sin\xi$ is not changing.

There is an integral of the BPS equations,
\begin{equation}
he^{-V}\,=\,k\sin\xi\,,
\end{equation}
where $k$ is a constant. We eliminate $h$ with the integral of motion and obtain
\begin{align}
f^{-1}\xi'\,=&\,-k^{-1}\left(s-B_z\right)e^{-V}\,, \notag \\
f^{-1}V'\,=&\,gW\sin\xi\,, \notag \\
f^{-1}u_i'\,=&\,-g\partial_{u_i}W\sin\xi\,, \notag \\
f^{-1}\phi'\,=&\,-\frac{g\partial_\phi{W}}{\sin\xi}\,,
\end{align}
with two constraints,
\begin{align} \label{constraint22}
\left(s-B_z\right)\,=&\,-k\left(2gWe^V\cos\xi+\kappa\right)\,, \notag \\
g\partial_\phi{W}\cos\xi\,=&\,\frac{1}{2}k^{-1}e^{-V}\partial_\phi{B}_z\,.
\end{align}

From the definition of $B_z$, we obtain
\begin{equation}
\partial_\phi{B}_z\,=\,e^{2\phi}D_z\sigma\,.
\end{equation}
For $\phi\ne0$, from the second equation in \eqref{constraint22}, we obtain
\begin{equation} \label{dzpsi}
D_z\sigma\,=\,\frac{2gke^V\partial_\phi{W}\cos\xi}{e^{2\phi}}\,,
\end{equation}
and the right hand side is independent of $\phi$. By differentiating \eqref{dzpsi}, we obtain fluxes expressed by
\begin{equation}
F^I_{yz}\,=\,\left(a^I\right)'\,=\,\left(\mathcal{I}^{(I)}\right)'\,,
\end{equation}
where we introduce
\begin{align}
\mathcal{I}^{(0)}\,\equiv&\,\frac{1}{\sqrt{2}}gke^V\cos\xi{e}^{-u_1-u_2-u_3}\,, \notag \\
\mathcal{I}^{(1)}\,\equiv&\,\frac{1}{\sqrt{2}}gke^V\cos\xi{e}^{-u_1+u_2+u_3}\,, \notag \\
\mathcal{I}^{(2)}\,\equiv&\,\frac{1}{\sqrt{2}}gke^V\cos\xi{e}^{u_1-u_2+u_3}\,, \notag \\
\mathcal{I}^{(3)}\,\equiv&\,\frac{1}{\sqrt{2}}gke^V\cos\xi{e}^{u_1+u_2-u_3}\,.
\end{align}

A symmetry of the BPS equations is
\begin{equation} \label{hsymm}
h\,\rightarrow\,-h\,, \qquad z\,\rightarrow\,-z\,,
\end{equation}
if $B_z\rightarrow-B_z$, $s\rightarrow-s$, $a^I\rightarrow-a^I$, $k\rightarrow-k$ and $F_{23}^I\rightarrow-F_{23}^I$. The frame is invariant under the transformation. By this symmetry we fix $h\le0$ in the main text.

\bibliographystyle{JHEP}
\bibliography{spindles.bib}

\end{document}